\documentclass[prc,letterpaper,showpacs,12pt,floatfix,preprint]{revtex4}

\usepackage{amsmath}
\usepackage{bm}
\usepackage{graphicx}

\begin{document}

\preprint{\vbox{\hbox{JLAB-THY-09-1036}}}

\title{Target Polarization for $^2 \vec H(e,e'p)n$ at GeV energies}

\author{Sabine Jeschonnek$^{(1)}$ and J. W. Van Orden$^{(2,3)}$}

\affiliation{\small \sl (1) The Ohio State University, Physics
Department, Lima, OH 45804\\
(2) Department of Physics, Old Dominion University, Norfolk, VA
23529\\and\\ (3) Jefferson Lab\footnote{Notice: Authored by Jefferson Science Associates, LLC under U.S. DOE Contract No. DE-AC05-06OR23177. The U.S. Government retains a non-exclusive, paid-up, irrevocable, world-wide license to publish or reproduce this manuscript for U.S. Government purposes.}, 12000 Jefferson Avenue, Newport
News, VA 23606
 }

\date{\today}

\begin{abstract}
We perform a fully relativistic calculation of the $^2 \vec H(e,e'p)n$ reaction in the impulse approximation employing the Gross equation
to describe the deuteron ground state, and we use the SAID parametrization of the full NN scattering amplitude
to describe the final state interactions (FSIs). The formalism for treating target polarization
with arbitrary polarization axes is discussed, and general properties of some asymmetries are derived from it.
We show results for momentum distributions and angular distributions of various asymmetries that can only be accessed with polarized targets.

\end{abstract}

\maketitle

\section{Introduction}

There are many interesting questions to be answered in investigating exclusive electron scattering from the deuteron:
what does the nuclear ground state look like at short distances, are there any six-quark contributions to the wave function,
when does a description in terms of hadronic degrees of freedom break down? In order to answer any of these questions,
a precise understanding of the reaction mechanism is mandatory. Final state interactions are the most relevant component
of the reaction mechanism at GeV energies, but meson exchange currents and isobar states will also contribute. The fact that
the deuteron is the simplest nucleus enables us to study all facets of the reaction mechanism in great detail. Anything
that can be gleaned from the deuteron will be highly useful for heavier nuclei. Exclusive electron scattering from nuclei
is one type of reaction where one may observe color transparency \cite{cteep}, and the deuteron itself provides a laboratory
for the study of neutrons, e.g. the neutron magnetic form factor \cite{hallbgmn}. The short range structures studied
in exclusive electron scattering might even reveal information about the properties of neutron stars \cite{misakneutstar}.
For some recent reviews of exclusive electron scattering, see e.g. \cite{wallyreview,ronfranz,sickreview}.

Recently \cite{bigdpaper}, we performed a fully relativistic calculation of the $D(e,e'p)n$ reaction, using
a relativistic wave function \cite{wallyfranzwf} and
$NN$ scattering data \cite{said} for our calculation of the full, spin-dependent final state interactions (FSIs).
The main difference to many other high quality calculations  using the generalized
eikonal approximation \cite{misak,ciofi,genteikonal} or a diagrammatic
approach \cite{laget} is the inclusion of all the
spin-dependent pieces in the nucleon-nucleon amplitude. Full FSIs have recently been included in \cite{schiavilla}.
 Several experiments with unpolarized deuterons
are currently under analysis or have been published recently, \cite{egiyansrc,halladata,hallbgmn,jerrygexp,blast}.
There are also new proposals for D(e,e'p) experiments at Jefferson Lab \cite{wernernewprop}.

In \cite{bigdpaper}, we focused on observables that are accessible for an unpolarized target and an unpolarized
nucleon detected in the final state. The spin-dependent pieces in our FSI calculation were particularly relevant for
the fifth response function, an observable that can be measured only with polarized electron beams.
Naturally, experiments with polarization of the target or ejectile are harder to perform than their unpolarized
counterparts. However, the extra effort allows one to study otherwise inaccessible observables that are rather
sensitive to certain properties of the nuclear ground state and the reaction mechanism.
In this paper, we investigate the asymmetries that can be measured with a polarized deuteron target.
These observables are of particular interest to us as we have a precise, fully spin-dependent description of the
final state interactions. As before, the focus of our
numerical calculations is the kinematic region accessible at GeV energies, i.e. the kinematic range of Jefferson Lab.
Currently, some deuteron target polarization data that were taken in Jefferson Lab's Hall B are being analyzed \cite{sebastian}.
At lower energies, measurements of the beam-vector asymmetry $A^V_{ed}$ have been performed at NIKHEF \cite{doug}
and at MIT Bates \cite{blast,blast2}. A formalism was developed within a non-relativistic framework, and
calculations of various asymmetries at lower energies were performed in \cite{arenhovel}. The tensor asymmetry has been
discussed at higher energies within a Glauber theory approach, with just a central FSI, in \cite{sabinetensor}.

The paper is organized as follows: first, we establish the general formalism necessary to calculate response functions
for polarized targets, and we discuss how to perform these calculations in two different coordinate systems. Then,
we continue with the calculation of asymmetries, and with the issues presented in using the experimental convention of
measuring polarizations along the electron beam direction versus the theoretical choice of polarization axis along the
three-momentum transfer $\vec q$. In the next section, we present our numerical results, in a kinematic region relevant to
experiments at Jefferson Lab. We show both momentum distributions and angular distributions, and we discuss the contributions
of the various spin-dependent parts of the final state interactions, as well as the influence of the ground state wave
function. We conclude with a brief summary.

\section{Formalism}

\subsection{Differential Cross Section}

\begin{figure}
\centerline{\includegraphics[height=3in]{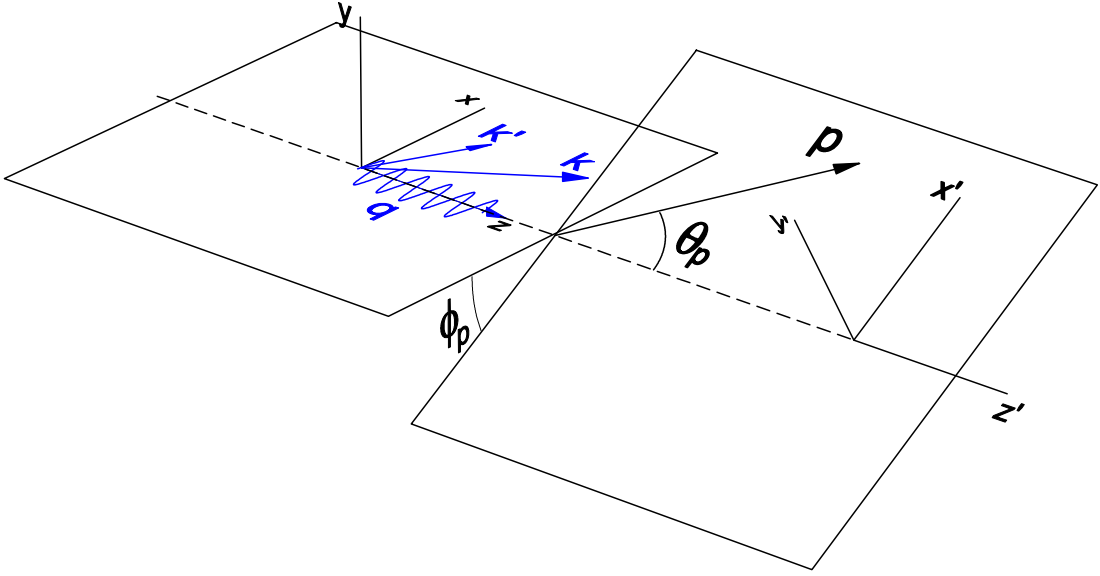}}
\caption{(Color online) Coordinate systems for the $D(e,e'p)$ reaction.  $k$ and $k'$ are the initial and final electron four-momenta,
$q$ is the four-momentum of the virtual photon and $p$ is the four-momentum of the final-state proton.  }\label{coordinates}
\end{figure}

The standard coordinate systems used to describe the $D(e,e'p)$ reaction are shown in Fig.\ref{coordinates}.
The initial and final electron momenta $\bm{k}$ and $\bm{k'}$ define the electron scattering plane
and the $xyz$-coordinate system is defined such that the $z$ axis, the quantization axis, lies along
the momentum of the virtual photon $q$ with the $x$-axis in the electron scattering plane and the $y$-axis
perpendicular to the plane. The momentum $p$ of the outgoing proton is in general not in this plane and is
located relative to the $xyz$ system by the polar angle $\theta_p$ and the azimuthal angle $\phi_p$.
A second coordinate system $x'y'z'$ is chosen such that the $z'$-axis is parallel to the $z$-axis
and the $x'$-axis lies in the plane formed by $\bm{p}$ and $\bm{q}$ and the $y'$-axis is normal to this plane.

The general form of the $D(e,e'p)$ cross section can be written in the lab frame as
\cite{raskintwd,dmtrgross}

\begin{eqnarray}
\left ( \frac{ d \sigma^5}{d \epsilon' d \Omega_e d \Omega_p} \right
)_h  & = & \frac{m_p \, m_n \, p_p}{8 \pi^3 \, M_d} \,
\sigma_{Mott} \,
f_{rec}^{-1} \,
 \Big[  v_L R_L +   v_T R_T
 + v_{TT} R_{TT} + v_{LT}R_{LT}
  \nonumber \\
& & +  h \,  v_{LT'} R_{LT'}+h\,v_{T'}R_{T'}
\Big] \, , \label{xsdef}
\end{eqnarray}
where $M_d$, $m_p$ and $m_n$  are the masses of the deuteron, proton and neutron,
 $p_p=p_1$ and $\Omega_p$
are the momentum and solid angle of the ejected proton, $\epsilon'$ is the
energy of the detected electron and $\Omega_e$ is its solid angle, with
$h=\pm 1$  for positive and negative electron helicity. The Mott cross
section is
\begin{equation}
\sigma_{Mott} = \left ( \frac{ \alpha \cos(\theta_e/2)} {2
\varepsilon \sin ^2(\theta_e/2)} \right )^2
\end{equation}
and the recoil factor is given by
\begin{equation}
f_{rec} = \left| 1+ \frac{\omega p_p - E_p q \cos \theta_p} {M_d \, p_p}
\right| \, . \label{defrecoil}
\end{equation}
The leptonic coefficients $v_K$ are
\begin{eqnarray}
v_L&=&\frac{Q^4}{q^4}\\
v_T&=&\frac{Q^2}{2q^2}+\tan^2\frac{\theta_e}{2}\\
v_{TT}&=&-\frac{Q^2}{2q^2}\\
v_{LT}&=&-\frac{Q^2}{\sqrt{2}q^2}\sqrt{\frac{Q^2}{q^2}+\tan^2\frac{\theta_e}{2}}\\
v_{LT'}&=&-\frac{Q^2}{\sqrt{2}q^2}\tan\frac{\theta_e}{2}\\
v_{T'}&=&\tan\frac{\theta_e}{2}\sqrt{\frac{Q^2}{q^2}+\tan^2\frac{\theta_e}{2}}
\end{eqnarray}
Within this general framework, we have two options for evaluating the response functions:
first, we will give expressions for the response functions in terms of matrix elements
that are defined with respect to the electron plane, i.e. the $xyz$ plane. These matrix elements
are implicitly dependent on $\phi_p$, the angle between hadron plane and electron plane, and these are the responses
used e.g. in \cite{raskintwd}.
Second, we give expressions for the responses in the $x'y'z'$ plane. All quantities given relative to the
$x'y'z'$ coordinate system are denoted by a line over the quantity. The current matrix elements, and therefore the
response functions, in the $x'y'z'$ coordinate system do not have any $\phi_p$ dependence. It is much more
practical to evaluate the responses in the $x'y'z'$ coordinate system. The commonly used responses
in the $xyz$ system can then easily be
obtained by accounting for the $\phi_p$ dependence explicitly, see eq.(\ref{connection}) below,
instead of newly evaluating matrix elements for each value of $\phi_p$.

Note that both coordinate systems use the same quantization axis: the $z$ axis and the $z'$ axis are parallel.
We will discuss using a different polarization along the beam, as commonly done by experimentalists, in the
next subsection on asymmetries.

The hadronic tensor for scattering from polarized deuterons is defined as

\begin{equation}
w_{\lambda'_\gamma,\lambda_\gamma}(D)=\sum_{s_1,s_2,\lambda_d,\lambda'_d}
\left<\bm{p}_1s_1;\bm{p}_2s_2;(-)\right|J_{\lambda'_\gamma}\left|\bm{P}\lambda'_d\right>^*
\left<\bm{p}_1s_1;\bm{p}_2s_2;(-)\right|J_{\lambda_\gamma}\left|\bm{P}\lambda_d\right>
\rho_{\lambda_d\lambda'_d}
\end{equation}
where
\begin{equation}
J_{\pm 1}=\mp\frac{1}{\sqrt{2}}(J^1\pm J^2)
\end{equation}
and
\begin{equation}
J_0=J^0
\end{equation}
is the charge operator. The notation $(-)$ in the final state indicates that the state satisfies the boundary conditions appropriate for an ``out'' state.
The deuteron density matrix in the $xyz$-frame is
\begin{equation}
\bm{\rho}=\frac{1}{3}\left(
\begin{array}{ccc}
1+\sqrt{\frac{3}{2}}\,T_{10}+\frac{1}{\sqrt{2}}T_{20}
& -\sqrt{\frac{3}{2}}(T^*_{11}+T^*_{21})
& \sqrt{3}\,T^*_{22}\\
-\sqrt{\frac{3}{2}}(T_{11}+T_{21})
& 1-\sqrt{2}\,T_{20}
& -\sqrt{\frac{3}{2}}(T^*_{11}-T^*_{21})\\
\sqrt{3}\,T^*_{22}
& -\sqrt{\frac{3}{2}}(T_{11}-T_{21})
& 1-\sqrt{\frac{3}{2}}\,T_{10}+\frac{1}{\sqrt{2}}T_{20}
\end{array}\right)
\end{equation}
and the set of tensor polarization coefficients is defined as
\begin{equation}
D=\left\{ U,T_{10},T_{11},T_{20},T_{21},T_{22}\right\}
\end{equation}
with $U$ designating the contribution from unpolarized deuterons. The derivation of the density matrix and the conventions used are described in the Appendix.

The response functions in the xyz-frame are given by
\begin{eqnarray}
R_L(D)&=&w_{00}(D)\nonumber\\
R_T(D)&=&w_{11}(D)+w_{-1-1}(D)\nonumber\\
R_{TT}(D)&=&2\Re(w_{1-1}(D))\nonumber\\
R_{LT}(D)&=&-2\Re(w_{01}(D)-w_{0-1}(D))\nonumber\\
R_{LT'}(D)&=&-2\Re(w_{01}(D)+w_{0-1}(D))\nonumber\\
R_{T'}(D)&=&w_{11}(D)-w_{-1-1}(D)
\label{respdefxyz}
\end{eqnarray}

Now we proceed to write down expressions for the responses in the $x'y'z'$ coordinate system. Calculating
the responses in this system offers a faster alternative to the above calculation, which requires
a new evaluation of the current matrix elements for each $\phi_p$ value.
The response functions defined above are implicitly dependent upon the angle $\phi_p$ between the electron plane and the hadron plane containing the proton and neutron in the final state.  This dependence can be made explicit by noting that
\begin{equation}
\left<\bm{p}_1s_1;\bm{p}_2s_2;(-)\right|J_{\lambda_\gamma}\left|\bm{P}\lambda_d\right>=e^{i(\lambda_d+\lambda_\gamma-s_1-s_2)\phi_p}
\overline{\left<\bm{p}_1s_1;\bm{p}_2s_2;(-)\right|J_{\lambda_\gamma}\left|\bm{P}\lambda_d\right>}
\end{equation}
where the line over the matrix elements is used to indicate that they are quantized relative to the $x'y'z'$ coordinate system. The hadronic tensor can then be written as
\begin{equation}
w_{\lambda'_\gamma,\lambda_\gamma}(D)=e^{-i(\lambda'_\gamma-\lambda_\gamma)\phi_p}\overline{w}_{\lambda'_\gamma,\lambda_\gamma}(\overline{D})
\label{connection}
\end{equation}
where
\begin{equation}
\overline{w}_{\lambda'_\gamma,\lambda_\gamma}(\overline{D})=\sum_{s_1,s_2,\lambda_d,\lambda'_d}
\overline{\left<\bm{p}_1s_1;\bm{p}_2s_2;(-)\right|J_{\lambda'_\gamma}\left|\bm{P}\lambda'_d\right>}^*
\overline{\left<\bm{p}_1s_1;\bm{p}_2s_2;(-)\right|J_{\lambda_\gamma}\left|\bm{P}\lambda_d\right>}
\overline{\rho}_{\lambda_d\lambda'_d}
\end{equation}
and
\begin{equation}
\overline{\rho}_{\lambda_d\lambda'_d}=e^{i(\lambda_d-\lambda'_d)\phi_p}\rho^D_{\lambda_d\lambda'_d}
\end{equation}
is the density matrix defined relative to the $x'y'z'$ coordinate system.

Using eq. (\ref{connection}) and the definition of the responses in the $xyz$ system, eq. (\ref{respdefxyz}),
the response functions in the $x'y'z'$ system then become
\begin{eqnarray}
R_L(\overline{D})&=&\overline{R}_L^{(I)}(\overline{D})\nonumber\\
R_T(\overline{D})&=&\overline{R}_T^{(I)}(\overline{D})\nonumber\\
R_{TT}(\overline{D})&=&\overline{R}_{TT}^{(I)}(\overline{D})\cos 2\phi_p+\overline{R}_{TT}^{(II)}(\overline{D})\sin 2\phi_p\nonumber\\
R_{LT}(\overline{D})&=&\overline{R}_{LT}^{(I)}(\overline{D})\cos\phi_p+\overline{R}_{LT}^{(II)}(\overline{D})\sin\phi_p\nonumber\\
R_{LT'}(\overline{D})&=&\overline{R}_{LT'}^{(I)}(\overline{D})\sin\phi_p+\overline{R}_{LT'}^{(II)}(\overline{D})\cos\phi_p\nonumber\\
R_{T'}(\overline{D})&=&\overline{R}_{T'}^{(II)}(\overline{D})\label{respdefbar}
\end{eqnarray}
where the reduced response functions for the two classes I and II are defined in terms of the hadronic tensors as
\begin{eqnarray}
\overline{R}^{(I)}_L(\overline{D})&=&\sum_i \overline{R}^{(I)}_L(\bm{\tau}^{(I)}_i)\overline{T}^{(I)}_i =  \overline{w}_{00}(\overline{D}) \nonumber \\
\overline{R}^{(I)}_T(\overline{D})&=&\sum_i \overline{R}^{(I)}_T(\bm{\tau}^{(I)}_i)\overline{T}^{(I)}_i
=\overline{w}_{1,1}(\overline{D})+\overline{w}_{-1,-1}(\overline{D})  \nonumber \\
\overline{R}^{(I)}_{TT}(\overline{D})&=&\sum_i \overline{R}^{(I)}_{TT}(\bm{\tau}^{(I)}_i)\overline{T}^{(I)}_i =2\Re(\overline{w}_{1,-1}(\overline{D}))  \nonumber \\
\overline{R}^{(II)}_{TT}(\overline{D})&=&\sum_i \overline{R}^{(II)}_{TT}(\bm{\tau}^{(II)}_i)\overline{T}^{(II)}_i = 2\Im(\overline{w}_{1,-1}(\overline{D}))\nonumber \\
\overline{R}^{(I)}_{LT}(\overline{D})&=&\sum_i \overline{R}^{(I)}_{LT}(\bm{\tau}^{(I)}_i)\overline{T}^{(I)}_i
=-2\Re(\overline{w}_{01}(\overline{D})-\overline{w}_{0-1}(\overline{D})) \nonumber   \\
\overline{R}^{(II)}_{LT}(\overline{D})&=&\sum_i \overline{R}^{(II)}_{LT}(\bm{\tau}^{(II)}_i)\overline{T}^{(II)}_i =2\Im(\overline{w}_{01}(\overline{D})+\overline{w}_{0-1}(\overline{D}))\nonumber\\
 \overline{R}^{(I)}_{LT'}(\overline{D})&=&\sum_i R^{(I)}_{LT'}(\bm{\tau}^{(I)}_i)\overline{T}^{(I)}_i =2\Im(\overline{w}_{01}(\overline{D})-\overline{w}_{0-1}(\overline{D})) \nonumber\\
R^{(II)}_{LT'}(\overline{D})&=&\sum_i \overline{R}^{(II)}_{LT'}(\bm{\tau}^{(II)}_i)\overline{T}^{(II)}_i = -2\Re(\overline{w}_{01}(\overline{D})+\overline{w}_{0-1}(\overline{D}))\nonumber\\
\overline{R}^{(II)}_{T'}(\overline{D})&=&\sum_i \overline{R}^{(II)}_{T'}(\bm{\tau}^{(II)}_i)\overline{T}^{(II)}_i = \overline{w}_{1,1}(\overline{D})-\overline{w}_{-1,-1}(\overline{D})\, , \label{defresp}
\end{eqnarray}
where
\begin{eqnarray}
&&\overline{T}^{(I)}_i \in \left\{ U,\Im(\overline{T}_{11}),\overline{T}_{20},\Re(\overline{T}_{21}),\Re(\overline{T}_{22})\right\}\nonumber\\
&&\overline{T}^{(II)}_i\in\left\{ \overline{T}_{10},\Re(\overline{T}_{11}),\Im(\overline{T}_{21}),\Im(\overline{T}_{22})\right\}
\label{deftau}
\end{eqnarray}
and
\begin{eqnarray}
&&\bm{\tau}^{(I)}_i \in \left\{ \bm{1},\bm{\tau}^\Im_{11},\bm{\tau}_{20},\bm{\tau}^\Re_{21},\bm{\tau}^\Re_{22}\right\}\nonumber\\
&&\bm{\tau}^{(II)}_i\in\left\{ \bm{\tau}_{10},\bm{\tau}^\Re_{11},\bm{\tau}^\Im_{21},\bm{\tau}^\Im_{22}\right\}\,.
\end{eqnarray}
The $\tau$-matrices are defined by (\ref{taumatrices}), (\ref{taureal}) and (\ref{tauimaginary}). The type $I$ and $II$ response functions can be obtained directly by noting that the density matrix can be written as
\begin{equation}
\overline{\bm{\rho}}=\frac{1}{3}\left(\bm{1}+\sum_i\bm{\tau}^{(I)}_i\overline{T}^{(I)}_i+\sum_i\bm{\tau}^{(II)}_i\overline{T}^{(II)}_i\right)\,.
\end{equation}
Defining a set of projected hadronic tensors as
\begin{equation}
\overline{w}_{\lambda'_\gamma,\lambda_\gamma}(\bm{\tau}^{(I,II)}_i)=\frac{1}{3}\sum_{s_1,s_2,\lambda_d,\lambda'_d}
\overline{\left<\bm{p}_1s_1;\bm{p}_2s_2;(-)\right|J_{\lambda'_\gamma}\left|\bm{P}\lambda'_d\right>}^*
\overline{\left<\bm{p}_1s_1;\bm{p}_2s_2;(-)\right|J_{\lambda_\gamma}\left|\bm{P}\lambda_d\right>}
(\bm{\tau}^{(I,II)}_i)_{\lambda_d\lambda'_d}\,,
\end{equation}
the type $I$ and $II$ response functions are then obtained by replacing the hadronic tensors on the right-hand side of
the expressions in (\ref{defresp}) with each appropriate projected hadronic tensor in turn. Note that the $\tau$ matrices satisfy
\begin{equation}
\left(\bm{\tau}_i^{(I)}\right)_{-\lambda-\lambda'}=(-1)^M\left(\bm{\tau}_i^{(I)}\right)_{\lambda\lambda'}
\end{equation}
and
\begin{equation}
\left(\bm{\tau}_i^{(II)}\right)_{-\lambda-\lambda'}=(-1)^{M+1}\left(\bm{\tau}_i^{(II)}\right)_{\lambda\lambda'}\,.
\end{equation}

\subsection{Symmetries of the Current Matrix Elements}
\label{sym_section}

The current matrix elements used here are defined in \cite{bigdpaper}. The matrix elements quantized in the hadron plane $x'y'z'$ can be shown to satisfy the symmetry
\begin{equation}
\overline{\left<\bm{p}_1s_1;\bm{p}_2s_2;(-)\right|J_{\lambda_\gamma}\left|\bm{P}\lambda_d\right>}=
(-1)^{\lambda_\gamma+\lambda_d-s_1-s_2}
\overline{\left<\bm{p}_1-s_1;\bm{p}_2-s_2;(-)\right|J_{\lambda_{-\gamma}}\left|\bm{P}-\lambda_d\right>}\label{spinflip}
\end{equation}
by starting with
\begin{equation}
i\Sigma_2\gamma^0u(\overline{\bm{p}},s)=(-1)^{\frac{1}{2}+s}u(\overline{\bm{p}},-s)
\end{equation}
which relies on the fact that the nucleon momenta have, by construction, no $y'$ component when quantized in the hadron plane.

Application of parity and time reversal to these matrix elements requires that
\begin{equation}
\left<\bm{p}_1s_1;\bm{p}_2s_2;(-)\right|J_{\lambda_\gamma}\left|\bm{P}\lambda_d\right>=
(-1)^{\lambda_\gamma+\lambda_d-s_1-s_2}
\left<\bm{P}-\lambda_d\right|J_{\lambda_{-\gamma}}\left|\bm{p}_1-s_1;\bm{p}_2-s_2;(+)\right>\,.
\end{equation}
Combining this with (\ref{spinflip}) gives
\begin{eqnarray}
\overline{\left<\bm{p}_1s_1;\bm{p}_2s_2;(-)\right|J_{\lambda_\gamma}\left|\bm{P}\lambda_d\right>}&=&
\overline{\left<\bm{P}\lambda_d\right|J_{\lambda_{\gamma}}\left|\bm{p}_1s_1;\bm{p}_2s_2;(+)\right>}\nonumber\\
&=&\overline{\left<\bm{p}_1s_1;\bm{p}_2s_2;(+)\right|J_{\lambda_\gamma}\left|\bm{P}\lambda_d\right>}^*\,.
\end{eqnarray}
In the plane-wave approximation there is no difference between the $(-)$ and $(+)$ boundary conditions. So in this approximation the current matrix elements are real.

\subsection{Asymmetries}

The simple form of (\ref{xsdef}) is due to the choice of quantization axis associated with the plane determined by
 the virtual photon momentum and the ejectile momentum.  In practice, the polarization coefficients are determined
 relative to a coordinate system fixed in the laboratory with the axis of quantization along the electron beam momentum.
 This can be easily accommodated by rotating the density matrix.  The relationship between the density matrix
 in the $x'y'z'$ coordinate system and the system with the quantization axis $z''$ along the electron
 momentum $\bm{k}$ and with $y''$ parallel to $y$ is
\begin{equation}
\overline{\rho}_{\lambda_d\lambda'_d}=\sum_{\Lambda\Lambda'}D^1_{\lambda_d\Lambda}(-\phi_p,\theta_{kq},0)D^1_{\lambda'_d\Lambda'}(-\phi_p,\theta_{kq},0)
\tilde{\rho}^D_{\Lambda\Lambda'}
\end{equation}
where the tilde denotes the density matrix for the $x''y''z''$ coordinate system and $\theta_{kq}$ is the angle between the beam momentum $\bm{k}$ and the momentum transfer $\bm{q}$. The polarization coefficients $\overline{T}_{JM}$ can be found as functions of the $\widetilde{T}_{JM}$ by using
\begin{eqnarray}
\overline{T}_{J0}&=&{\rm Tr}(\bm{\tau}^\dag_{J0}\overline{\bm{\rho}})\nonumber\\
\Re(\overline{T}_{JM})&=&\frac{1}{2}{\rm Tr}\left[{\bm{\tau}^\Re_{JM}}^\dag \overline{\bm{\rho}}\right]\nonumber\\
\Im(\overline{T}_{JM})&=&\frac{1}{2}{\rm Tr}\left[{\bm{\tau}^\Im_{JM}}^\dag \overline{\bm{\rho}}\right]
\end{eqnarray}
The response functions for the $x''y''z''$ coordinate system can by found by using these in (\ref{defresp}).

The asymmetries that we will calculate here involve the case where $\widetilde{T}_{10}$ is nonzero
with all other polarization coefficients equal to zero, or where $\widetilde{T}_{20}$ is nonzero
with all other polarization coefficients equal to zero. In the first case,
\begin{eqnarray}
\overline{T}_{10}&=&\cos\theta_{kq}\tilde{T}_{10}\nonumber\\
\Re(\overline{T}_{11})&=&-\frac{1}{\sqrt{2}}\sin\theta_{kq}\cos\phi_p\tilde{T}_{10}\nonumber\\
\Im(\overline{T}_{11})&=&\frac{1}{\sqrt{2}}\sin\theta_{kq}\sin\phi_p\tilde{T}_{10}\nonumber\\
\overline{T}_{2M}&=&0\label{polT10}
\end{eqnarray}
while in the second case
\begin{eqnarray}
\overline{T}_{1M}&=&0\nonumber\\
\overline{T}_{20}&=&\frac{1}{4}(1+3\cos 2\theta_{kq})\tilde{T}_{20}\nonumber\\
\Re(\overline{T}_{21})&=&-\sqrt{\frac{3}{8}}\sin 2\theta_{kq}\cos\phi_p\tilde{T}_{20}\nonumber\\
\Im(\overline{T}_{21})&=&\sqrt{\frac{3}{8}}\sin 2\theta_{kq}\sin\phi_p\tilde{T}_{20}\nonumber\\
\Re(\overline{T}_{22})&=&\sqrt{\frac{3}{32}}(1-\cos 2\theta_{kq})\cos 2\phi_p\tilde{T}_{20}\nonumber\\
\Im(\overline{T}_{22})&=&-\sqrt{\frac{3}{32}}(1-\cos 2\theta_{kq})\sin 2\phi_p\tilde{T}_{20}\,.\label{polT20}
\end{eqnarray}

A similar relation between the $xyz$ and $x''y''z''$ coordinates systems is given by
\begin{equation}
\rho_{\lambda_d\lambda'_d}=\sum_{\Lambda\Lambda'}d^1_{\lambda_d\Lambda}(\theta_{kq})d^1_{\lambda'_d\Lambda'}(\theta_{kq})
\tilde{\rho}^D_{\Lambda\Lambda'}\,.
\end{equation}
Then,
\begin{equation}
T_{JM}=T_{JM}(\widetilde{D})\,.
\end{equation}
The relations between the polarization coefficients can be obtained from (\ref{polT10}) and (\ref{polT20}) by setting $\phi_p=0$ and making the replacements $\overline{T}_{JM}\rightarrow T_{JM}$.

The single and double asymmetries for these two polarizations are defined as
\begin{eqnarray}
\label{asymdef}
A^V_d&=&\frac{v_L R_L(\widetilde{T}_{10})+v_T R_T(\widetilde{T}_{10})+v_{TT}R_{TT}(\widetilde{T}_{10})+v_{LT} R_{LT}(\widetilde{T}_{10})}{\widetilde{T}_{10}\Sigma}\nonumber\\
A^T_d&=&\frac{v_L R_L(\widetilde{T}_{20})+v_T R_T(\widetilde{T}_{20})+v_{TT}R_{TT}(\widetilde{T}_{20})+v_{LT} R_{LT}(\widetilde{T}_{20})}{\widetilde{T}_{20}\Sigma}\nonumber\\
A^V_{ed}&=&\frac{v_{LT'}R_{LT'}(\widetilde{T}_{10})
+v_{T'}R_{T'}(\widetilde{T}_{10})}{\widetilde{T}_{10}\Sigma}\nonumber\\
A^T_{ed}&=&\frac{v_{LT'}R_{LT'}(\widetilde{T}_{20})+v_{T'}R_{T'}(\widetilde{T}_{20})
}{\widetilde{T}_{20}\Sigma}
\end{eqnarray}
where
\begin{equation}
\Sigma=v_L R_L(U)+v_T R_T(U)+v_{TT}R_{TT}(U)+v_{LT} R_{LT}(U)\,.
\end{equation}
 Here $R_i(\widetilde{T}_{10})$ and $R_i(\widetilde{T}_{20})$ denote the response functions where only $\widetilde{T}_{10}$ is nonzero or only $\widetilde{T}_{20}$ is nonzero. $R_i(U)$ denotes the unpolarized response functions.

Using the definitions of the asymmetries, the expressions for the $\bar{T}_{JM}$ as a function of the $\tilde T$
and the definitions of the response functions in the $x'y'z'$ system, one obtains the following symmetry relations
with respect to $\phi_p$:

\begin{eqnarray}
A^V_d (\phi_p) &=&  - A^V_d (360^o - \phi_p)  \nonumber \\
A^T_d (\phi_p) &=&  A^T_d (360^o - \phi_p)\nonumber \\
A^V_{ed} (\phi_p) &=&  A^V_{ed} (360^o - \phi_p)\nonumber \\
A^T_{ed} (\phi_p) &=& - A^T_{ed} (360^o - \phi_p)
\label{phisym}
\end{eqnarray}

\section{Results}

All results are shown for a quantization axis along the beam direction, as measured in experiments, not along the
direction of the three-momentum transfer.

\subsubsection{Momentum Distributions}

\begin{figure}[ht]
\includegraphics[width=14pc,angle=270]{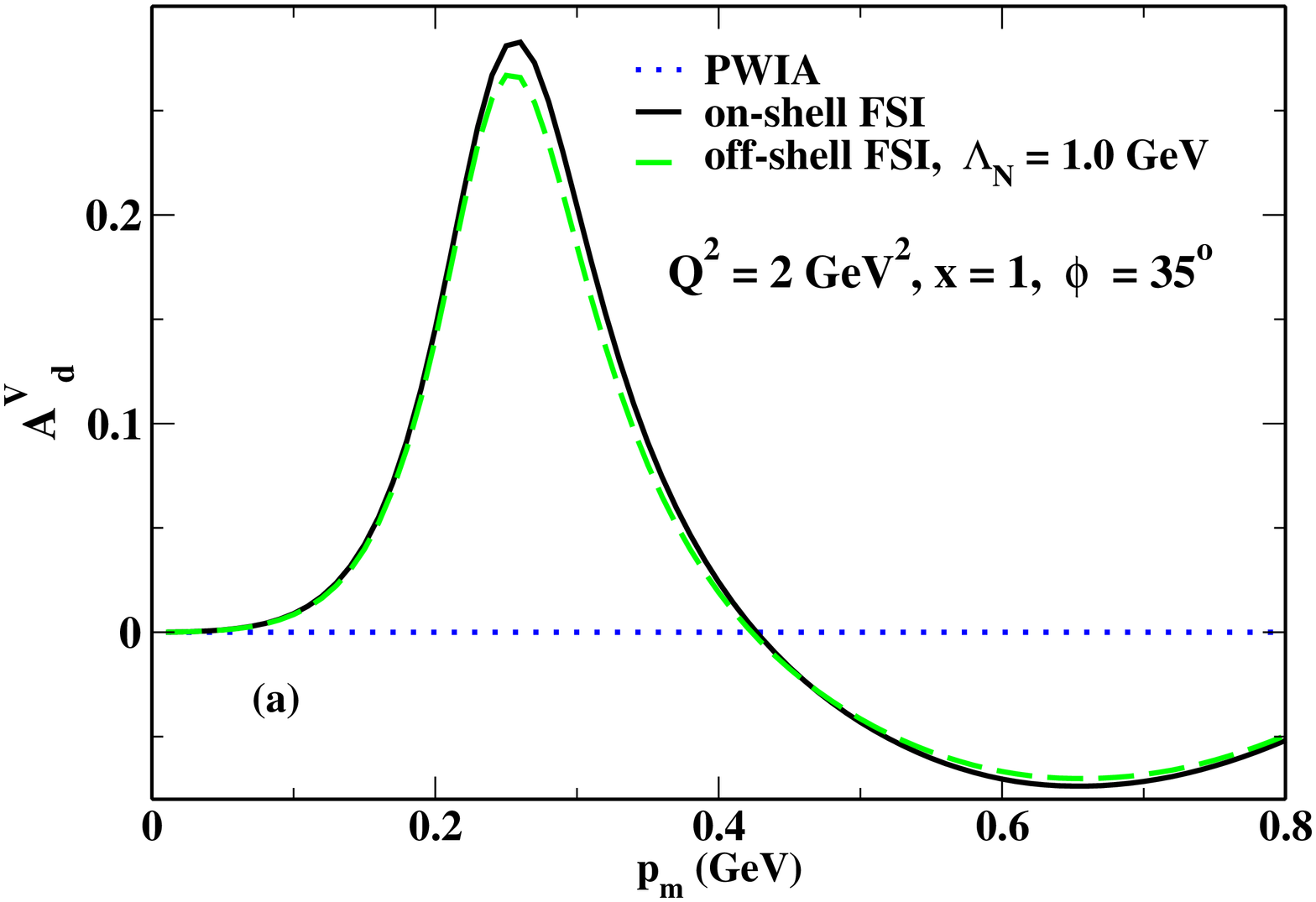}
\includegraphics[width=14pc,angle=270]{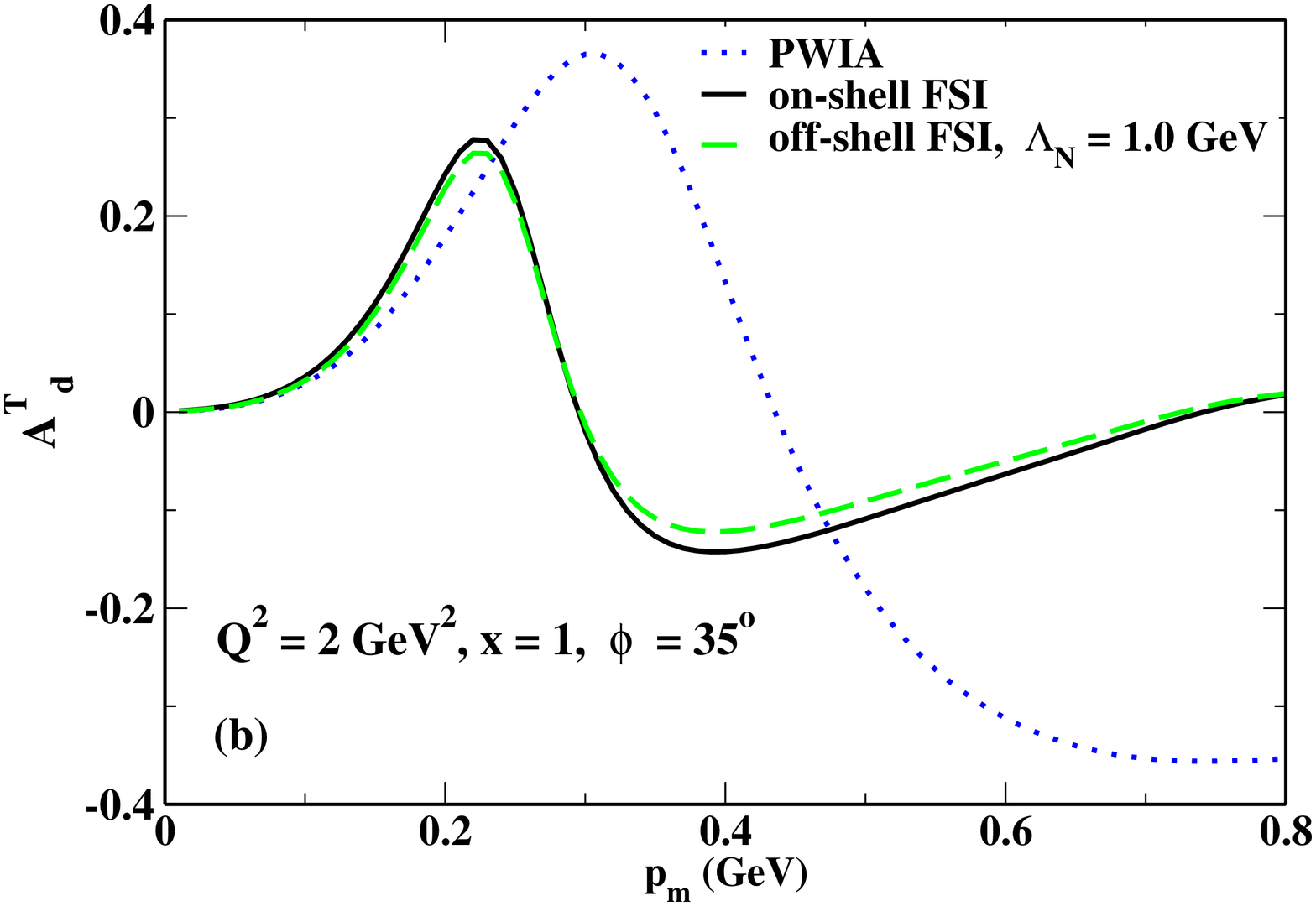}
\includegraphics[width=14pc,angle=270]{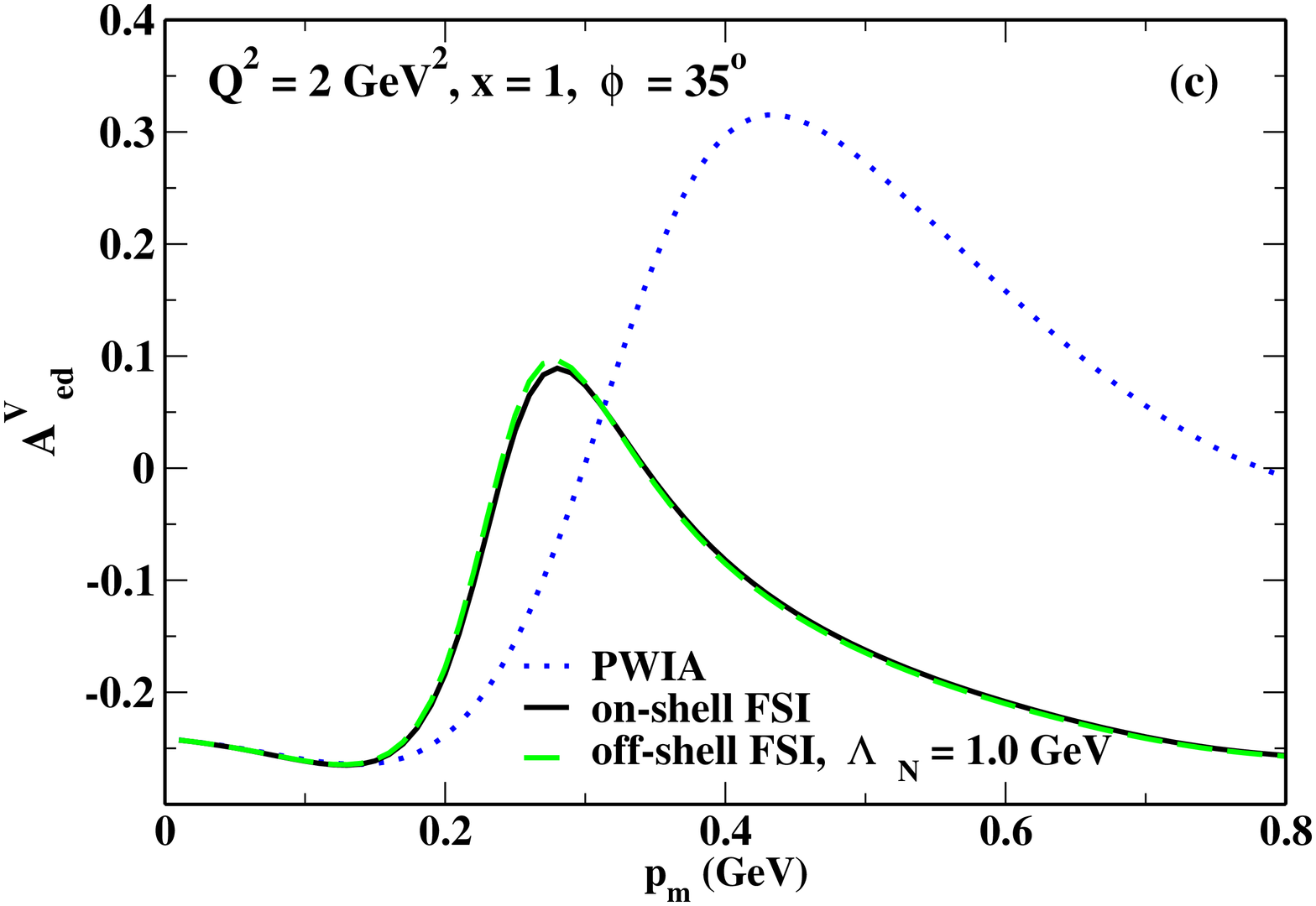}
\includegraphics[width=14pc,angle=270]{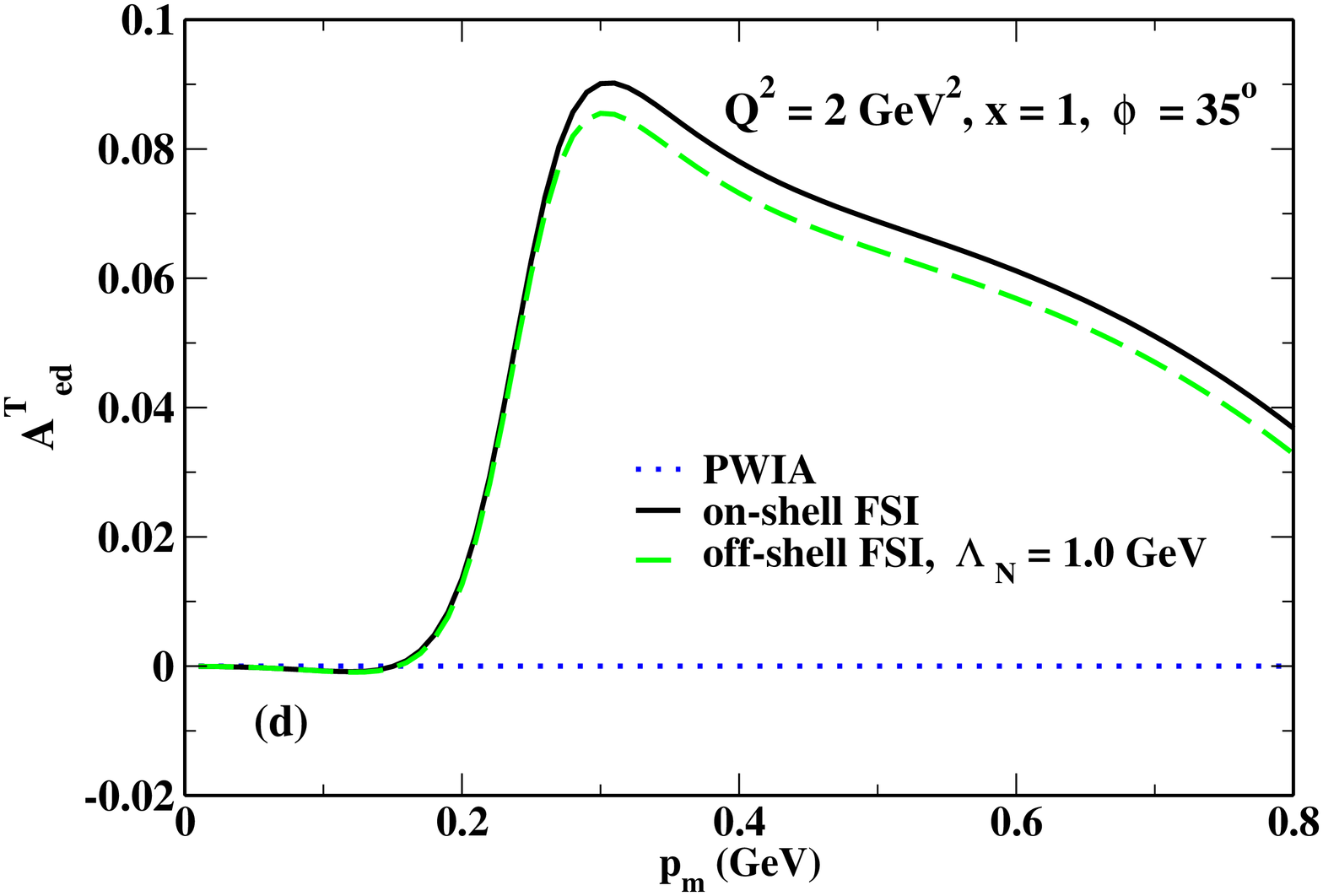}
\caption{(Color online) The asymmetries $A^V_d$ (panel (a)), $A^T_d$ (panel (b)), $A^V_{ed}$ (panel (c)) and $A^T_{ed}$ (panel (d))
for a beam energy of $5.5~
{\rm GeV}$, $Q^2 = 2 ~{\rm GeV}^2$, $x = 1$ GeV, and $\phi_p = 35^\circ$ are shown
calculated in PWIA (dotted line), with on-shell FSI (solid line), and with
on-shell and off-shell FSI (dashed line),
as a function of the missing momentum.
 }
\label{fig_md_x1_pwiafsi}
\end{figure}

In Fig. \ref{fig_md_x1_pwiafsi}, we show the four asymmetries for a four-momentum transfer of $Q^2 = 2$ GeV$^2$ and $x = 1$.
These kinematics correspond to quasi-elastic scattering. Note that in the plane-wave approximation the asymmetries $A^V_d$ and $A^T_{ed}$ vanish.  They are non-zero only when the FSIs are included.
This is predicted in non-relativistic PWIA calculations, and our relativistic approach does not change this feature.
In the cases where the asymmetries are non-zero for PWIA, the inclusion
of FSIs leads to
a shift, and a slight distortion, of the features that are already present in the asymmetries.
The dips and bumps become narrower
when FSIs are included, and they appear at somewhat lower missing momenta.
The difference between just on-shell FSI and full FSIs including on-shell and off-shell distributions
is very small. The largest off-shell FSI effects are present for larger missing momenta in $A^T_{ed}$.

From our discussion in section \ref{sym_section}, we can now explain the observed behavior of the asymmetries in PWIA:
all PWIA current matrix elements are real, and so any response that consists of taking the imaginary part of any part of the
hadronic tensor will vanish in PWIA. When consulting eq. (\ref{deftau}), we see that the vector asymmetries with $\bar{T}_{10} \not= 0$
are associated with the class II responses, and that the tensor asymmetries with $\bar{T}_{20} \not= 0$
are associated with the class I responses. From its definition, we can see that $A^V_d$ is associated with the vector, i.e. class II,
contributions to the L, T, TT, and LT responses. The L and T responses have no class II versions, and the class II versions
of the TT and LT responses are proportional to the imaginary part of certain pieces of the hadronic tensor. Thus, $A^V_d$ vanishes
in PWIA. A similar argument shows that $A^T_{ed}$ must vanish in PWIA, whereas the other two asymmetries will always have non-zero
contributions. This argument was made in the hadron plane, in the $x'y'z'$ frame. It is also valid when the quantization axis is rotated,
as the rotation itself will not lead to a non-zero value for an asymmetry that vanishes for one set of quantization axes.

\begin{figure}[ht]
\includegraphics[width=14pc,angle=270]{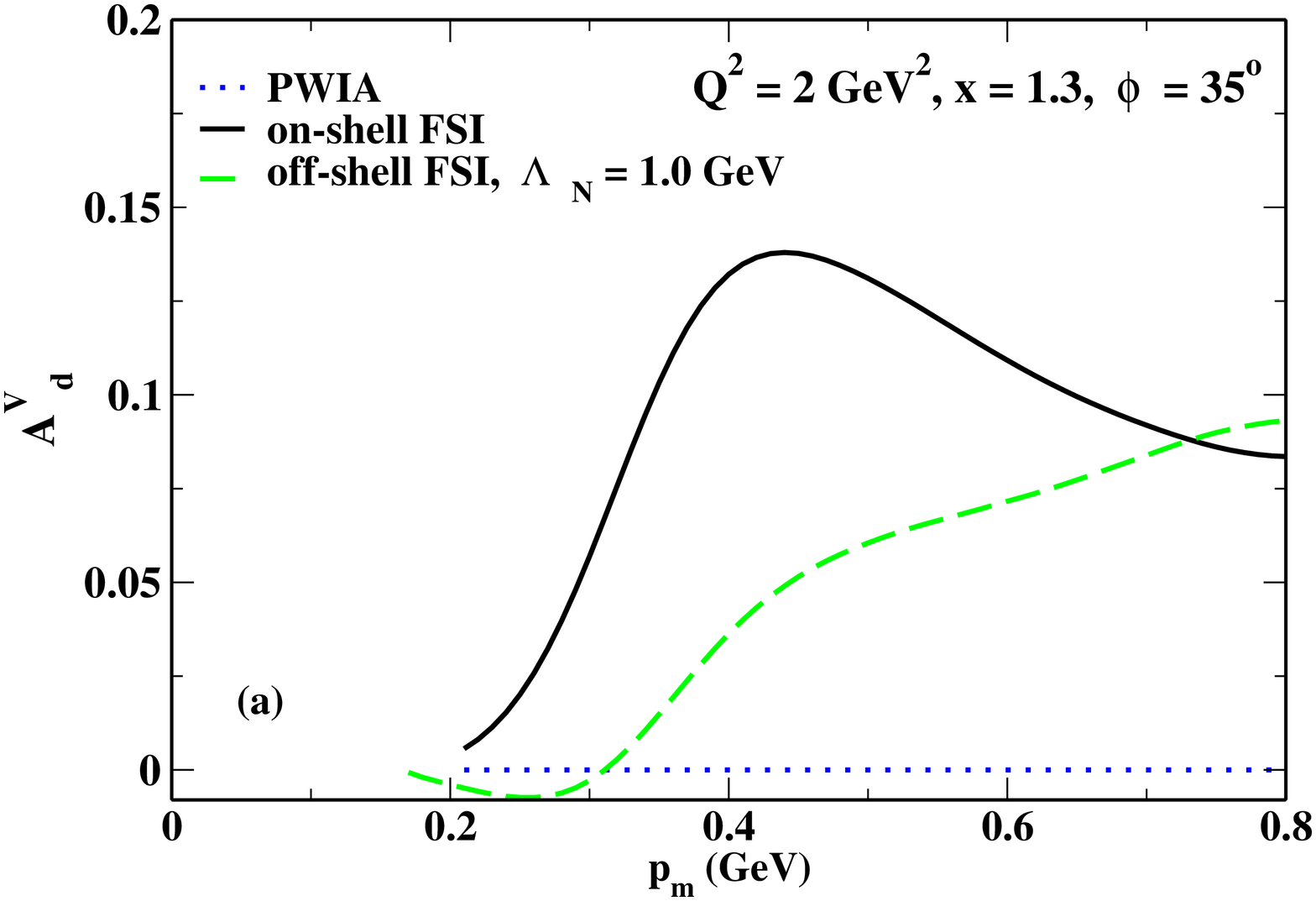}
\includegraphics[width=14pc,angle=270]{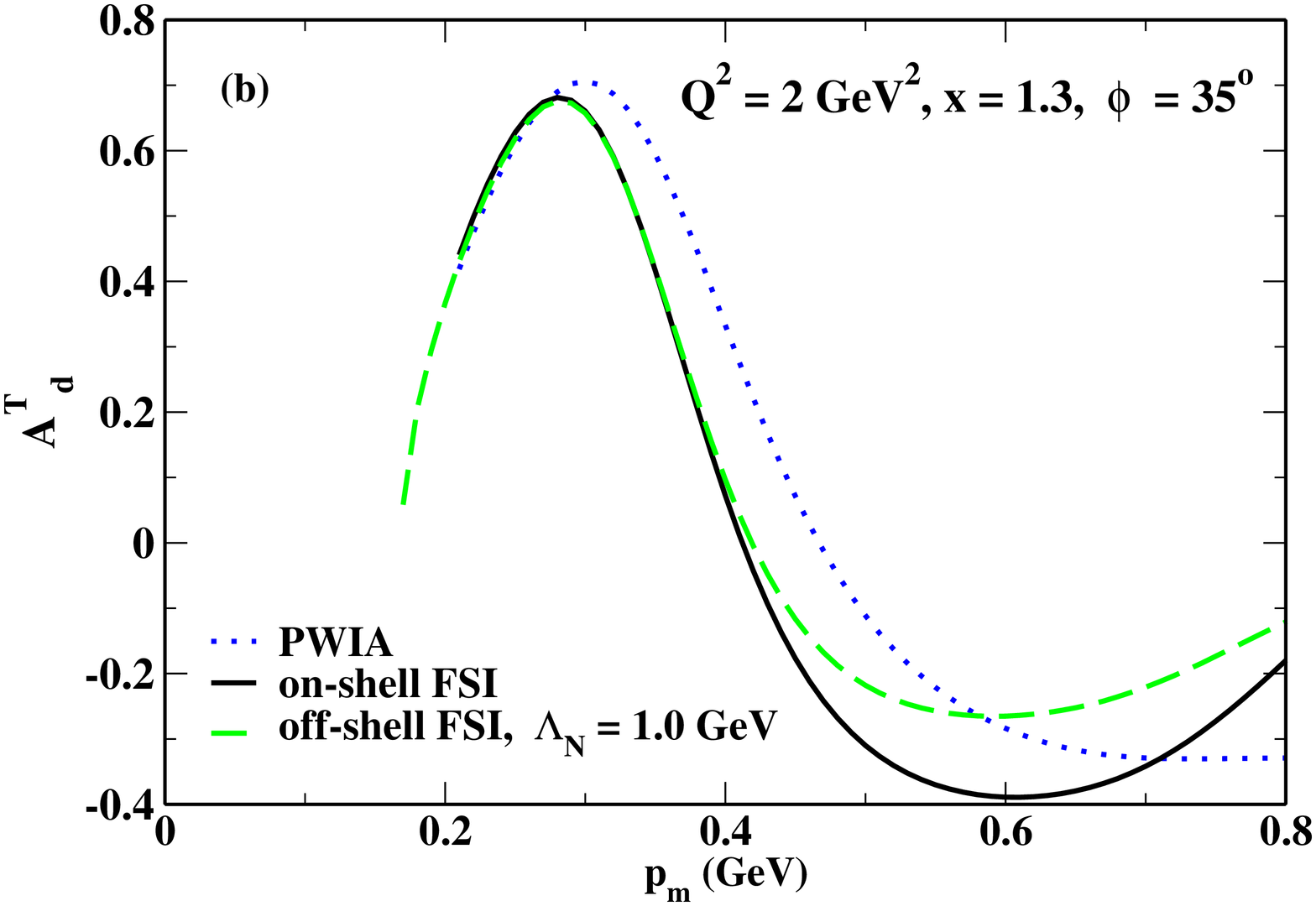}
\includegraphics[width=14pc,angle=270]{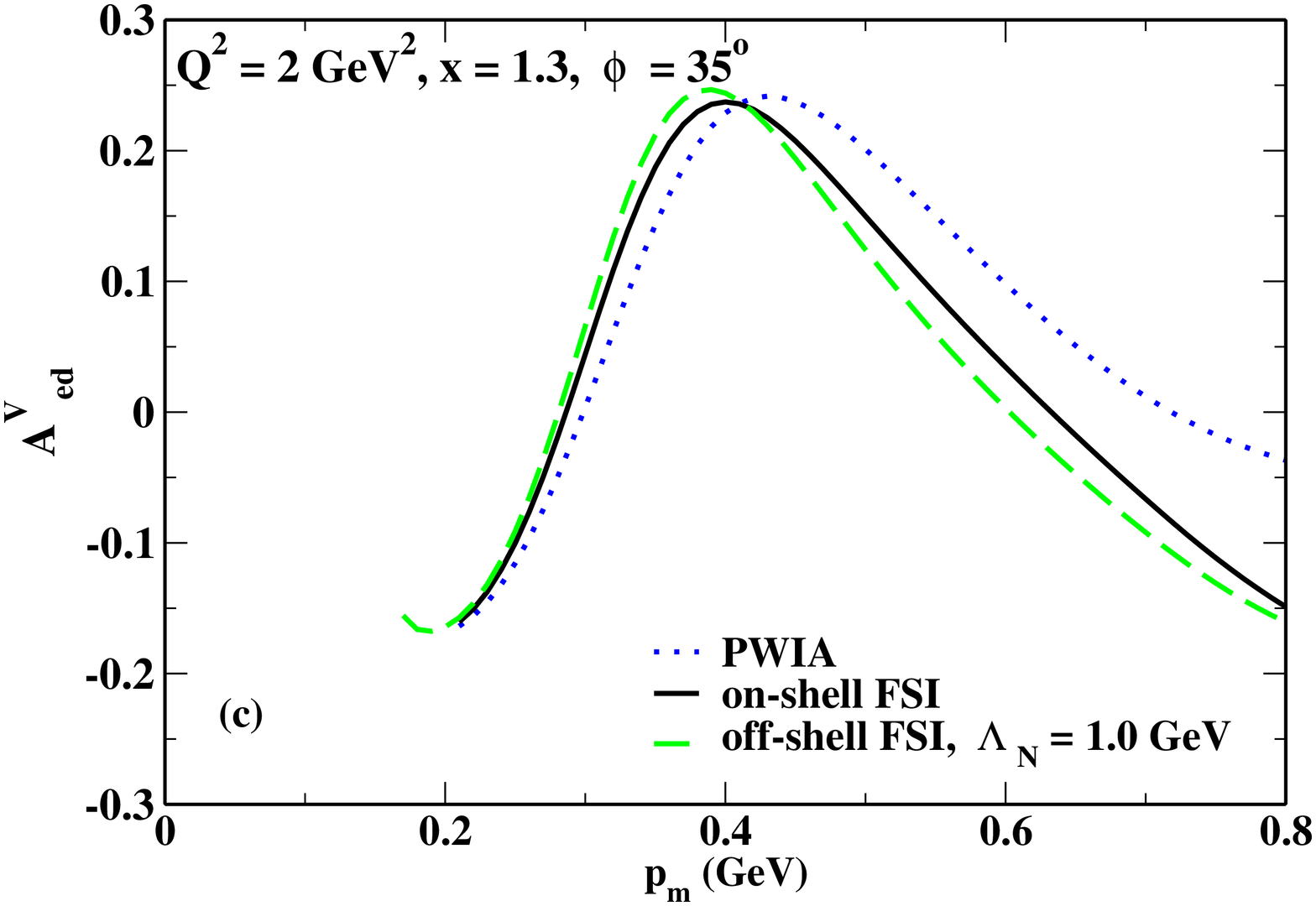}
\includegraphics[width=14pc,angle=270]{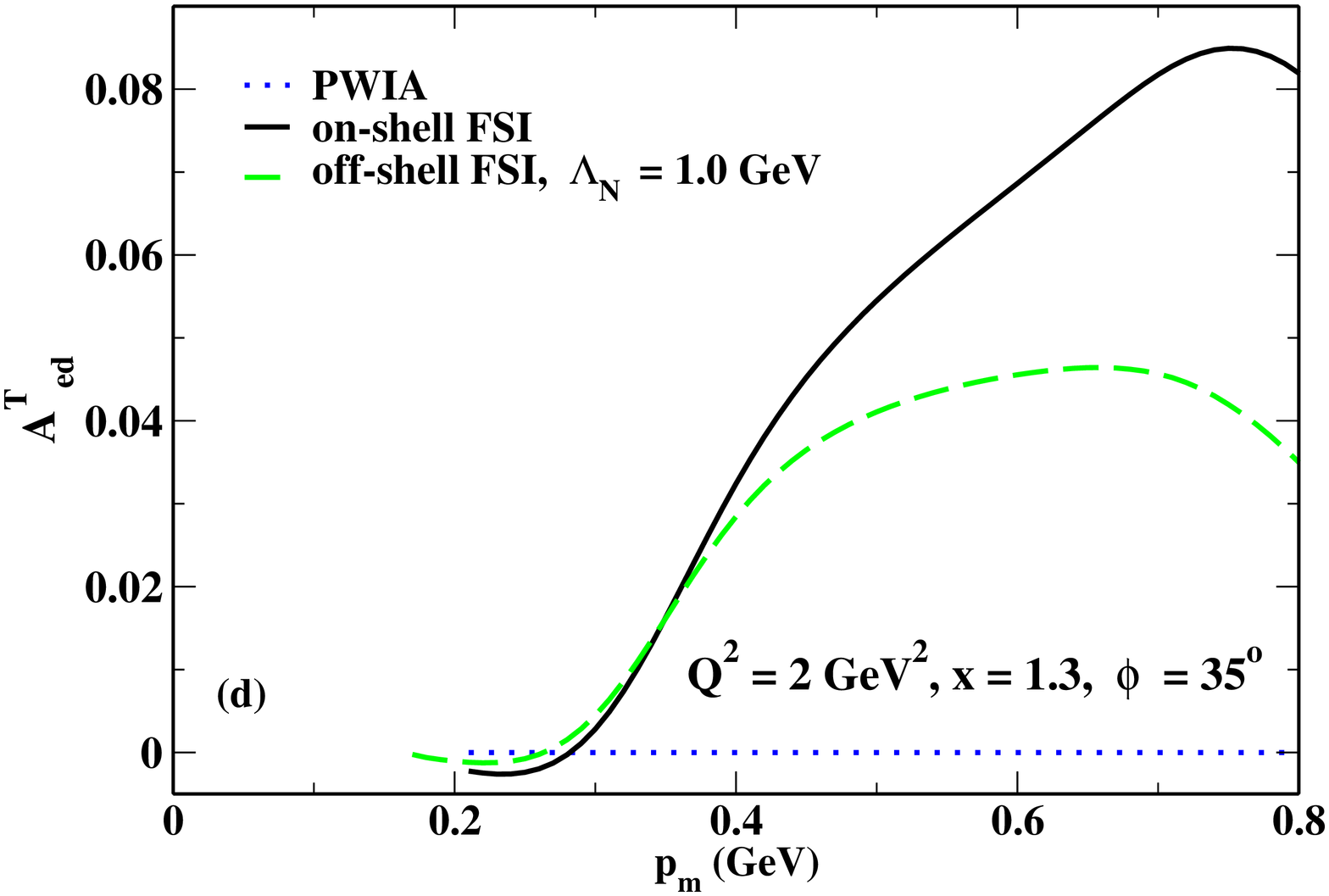}
\caption{(Color online) The asymmetries $A^V_d$ (panel (a)), $A^T_d$ (panel (b)), $A^V_{ed}$ (panel (c)) and $A^T_{ed}$ (panel (d))
for a beam energy of $5.5~
{\rm GeV}$, $Q^2 = 2 ~{\rm GeV}^2$, $x = 1.3$ GeV, and $\phi_p = 35^\circ$ are shown
calculated in PWIA (dotted line), with on-shell FSI (solid line), and with
on-shell and off-shell FSI (dashed line),
 as a function of the missing momentum. }
\label{fig_md_x13_pwiafsi}
\end{figure}

In Fig. \ref{fig_md_x13_pwiafsi}, we show the four asymmetries for a four-momentum transfer of $Q^2 = 2$ GeV$^2$ and $x = 1.3$.
These kinematics are away from the quasi-elastic peak, and we expect off-shell contributions to the FSIs to be more
relevant here. We have observed the increase in relative importance of the off-shell FSIs already for unpolarized
observables in \cite{bigdpaper}.
Due to the chosen kinematics, smaller values of the missing momentum are not accessible. As for the quasi-elastic kinematics shown above,
$A^V_d$ and $A^T_{ed}$, are non-zero only once FSIs are included, and the FSIs shift the bumps and dips to lower missing momenta.
The shift to lower momenta is much smaller here than for the quasi-elastic case, though.
In contrast to the $x = 1$ kinematics, the off-shell FSIs now play a more prominent role. The differences between just on-shell FSIs and
off-shell and on-shell FSIs are large for $A^V_d$ and $A^T_{ed}$, and they are apparent already at low missing momentum.
For the two other asymmetries, $A^V_{ed}$ and $A^T_{d}$, the differences are less pronounced and are most significant at the largest
missing momenta considered here. Having a non-zero asymmetry already in PWIA makes the asymmetry less sensitive to
off-shell effects: if the PWIA results are non-zero, the FSIs are very relevant corrections,
and the off-shell FSIs are less significant
corrections of the correction; if the PWIA results are zero, the FSIs provide the entire asymmetry,
and the off-shell FSI corrections are relevant.

\begin{figure}[ht]
\includegraphics[width=19pc,angle=0]{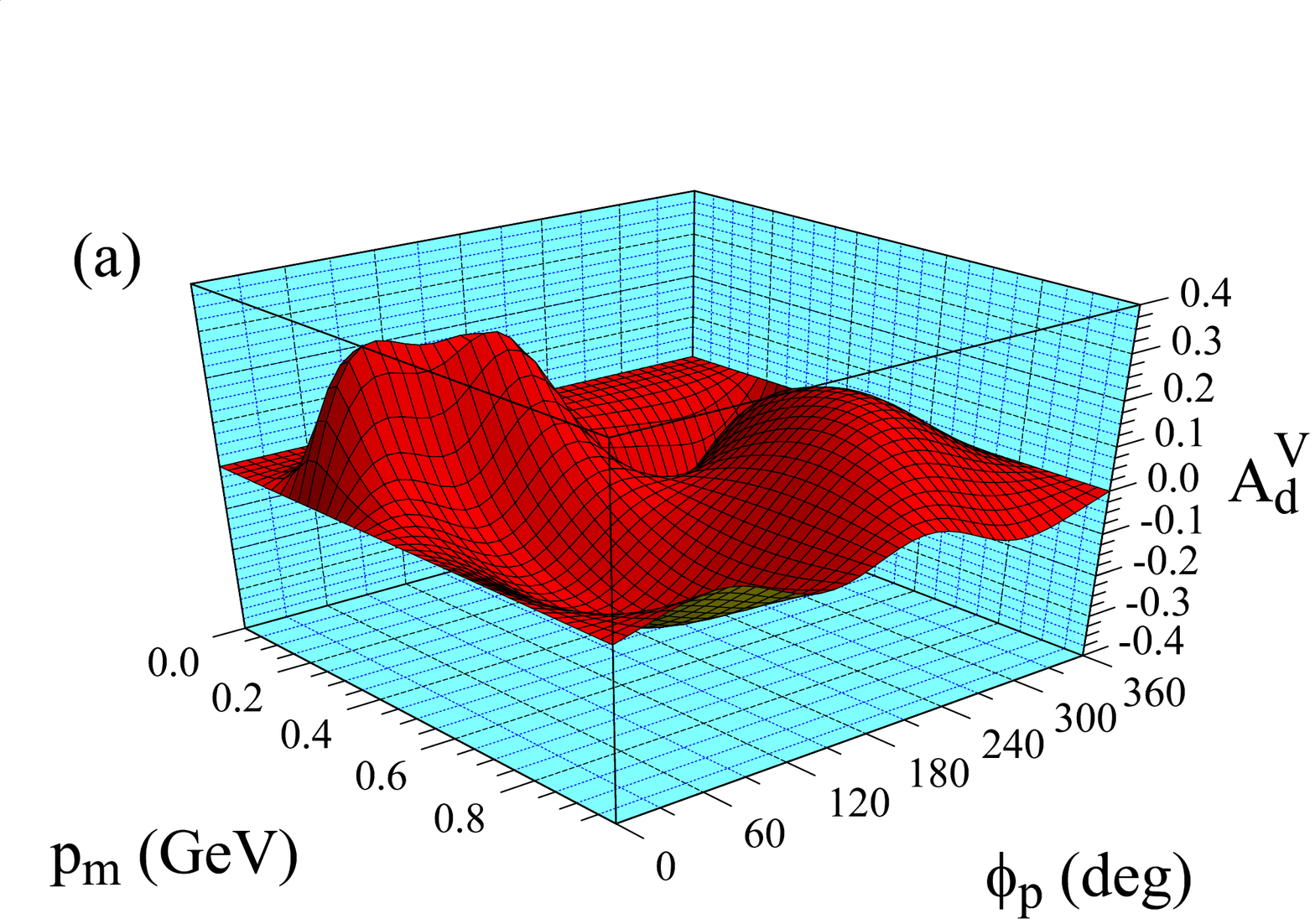}
\includegraphics[width=19pc,angle=0]{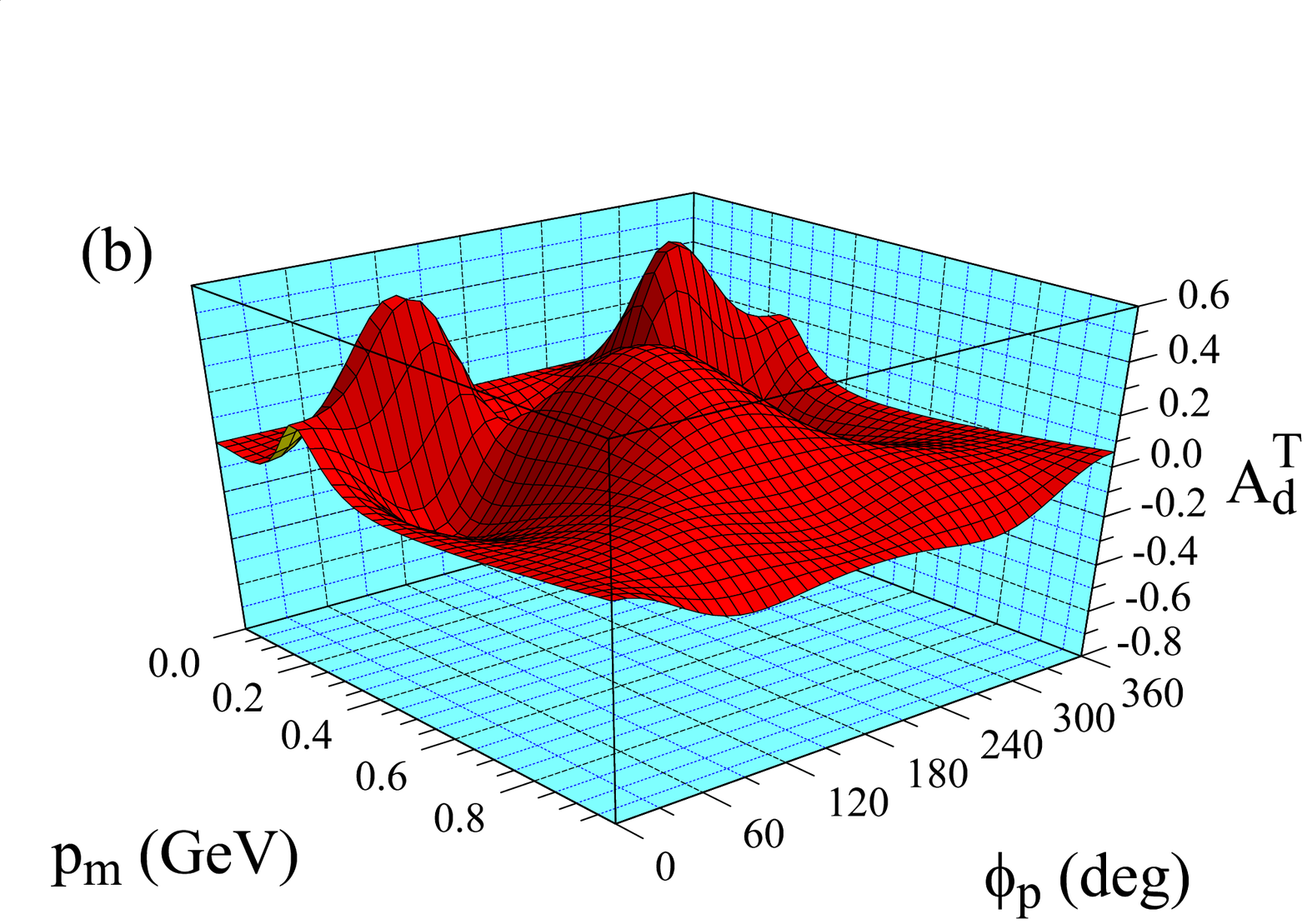}
\includegraphics[width=19pc,angle=0]{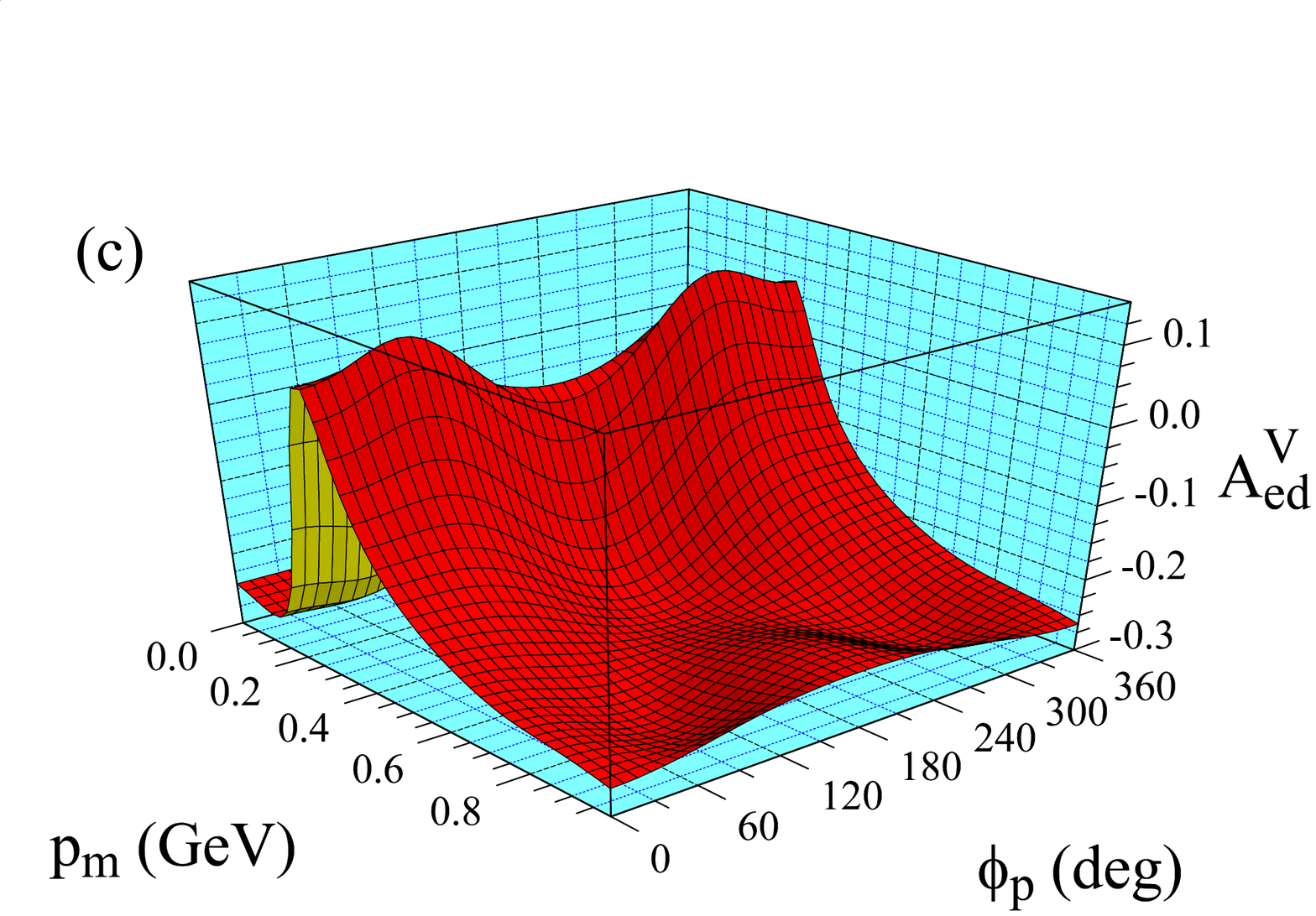}
\includegraphics[width=19pc,angle=0]{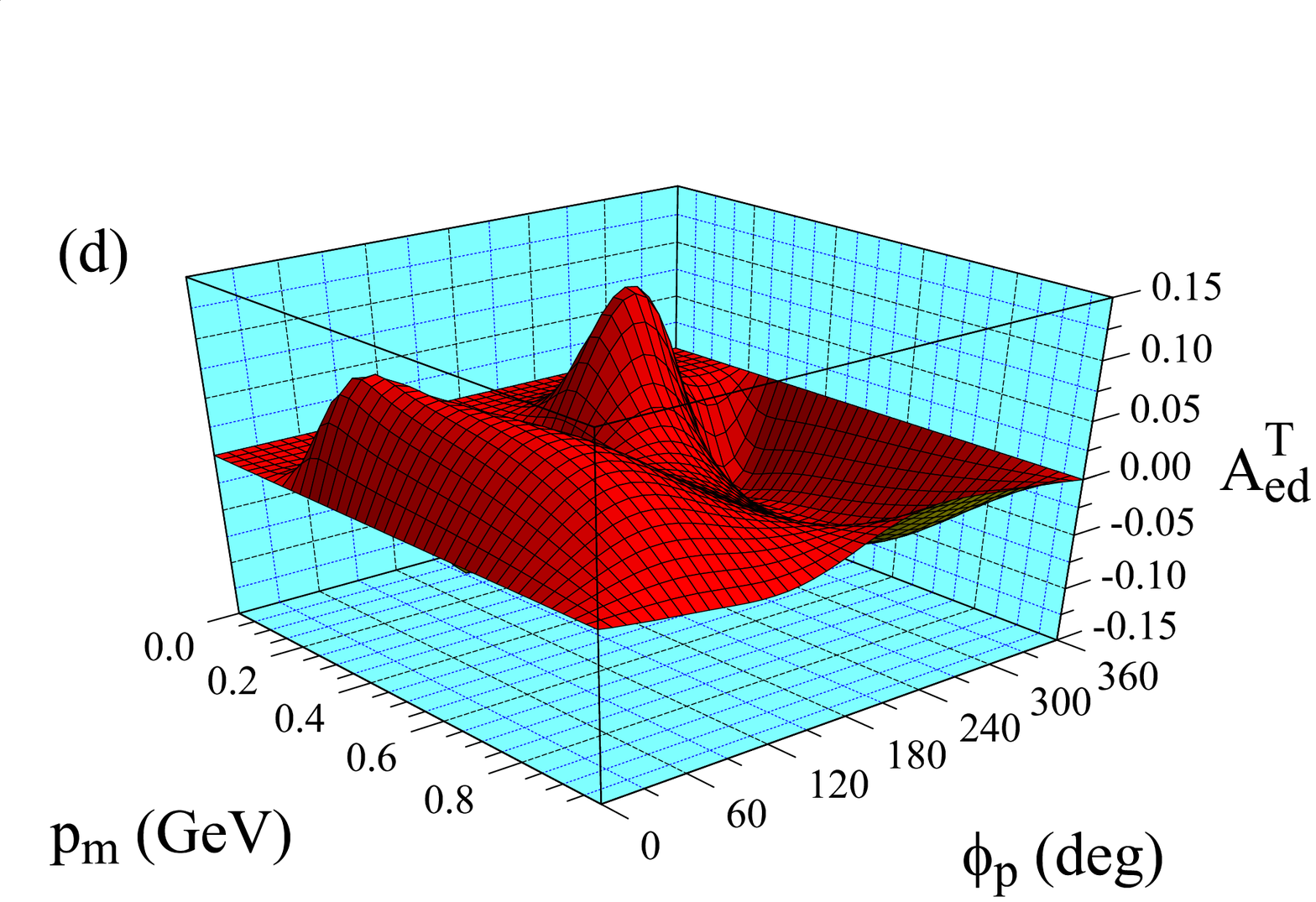}
\caption{(Color online) The asymmetries $A^V_d$ (a), $A^T_d$ (b), $A^V_{ed}$ (c), and $A^T_{ed}$ (d),
for a beam energy of $5.5~
{\rm GeV}$, $Q^2 = 2 ~{\rm GeV}^2$, and $x = 1$ are shown
calculated with on-shell FSI
 as a function of the missing momentum and the proton's azimuthal angle. }
\label{fig_3d_multi_md_qsq2_x1_fsi}
\end{figure}

The asymmetries that we investigate here also have a dependence on the azimuthal angle $\phi_p$ of the outgoing proton.
The two sets of figures above were shown for a value of $\phi_p = 35^o$. This value was chosen to avoid any special cases
for $\phi_p = 0^o, 45^o$, or $90^o$. However, the overall $\phi_p$ dependence is interesting, too.
We show this dependence for
all four asymmetries in Fig. \ref{fig_3d_multi_md_qsq2_x1_fsi} in a three-dimensional plot.

One can see that for $A^V_d$, the broad bump and dip structures observed for $\phi_p = 35^o$
turn into a broad dip and bump for
$\phi_p$ values above $180^o$, inverting the original, low $\phi_p$ structure.
A very similar inversion of the structures
is observed for $A^T_{ed}$: the broad ridge at lower $\phi_p$ turns into a valley for large $\phi_p$, and the sharp
dip at low missing momenta and medium $\phi_p$ turns into a peak at $\phi_p > 180^o$
For the other two asymmetries, $A^T_d$ and $A^V_{ed}$, the plots are
symmetric around $\phi_p = 180^o$. This is the behavior predicted by eq.(\ref{phisym}) for the four asymmetries.

For the kinematics away from the quasi-elastic peak, for $x = 1.3$,
the same type
of $\phi_p$ dependence and the same $\phi_p$ symmetries are observed, and we therefore do not display a separate figure.
The asymmetries reach much larger maximum values for $x = 1.3$, though.

\subsubsection{Angular Distributions}

We now discuss our results for angular distributions. Note that for the FSI calculations, there is a limit to the kinematic region
we can calculate for, as the proton-neutron scattering amplitude that we use is available only for $pn$ energies up to $1.3$~GeV,
see \cite{bigdpaper} for details.

\begin{figure}[ht]
\includegraphics[width=14pc,angle=270]{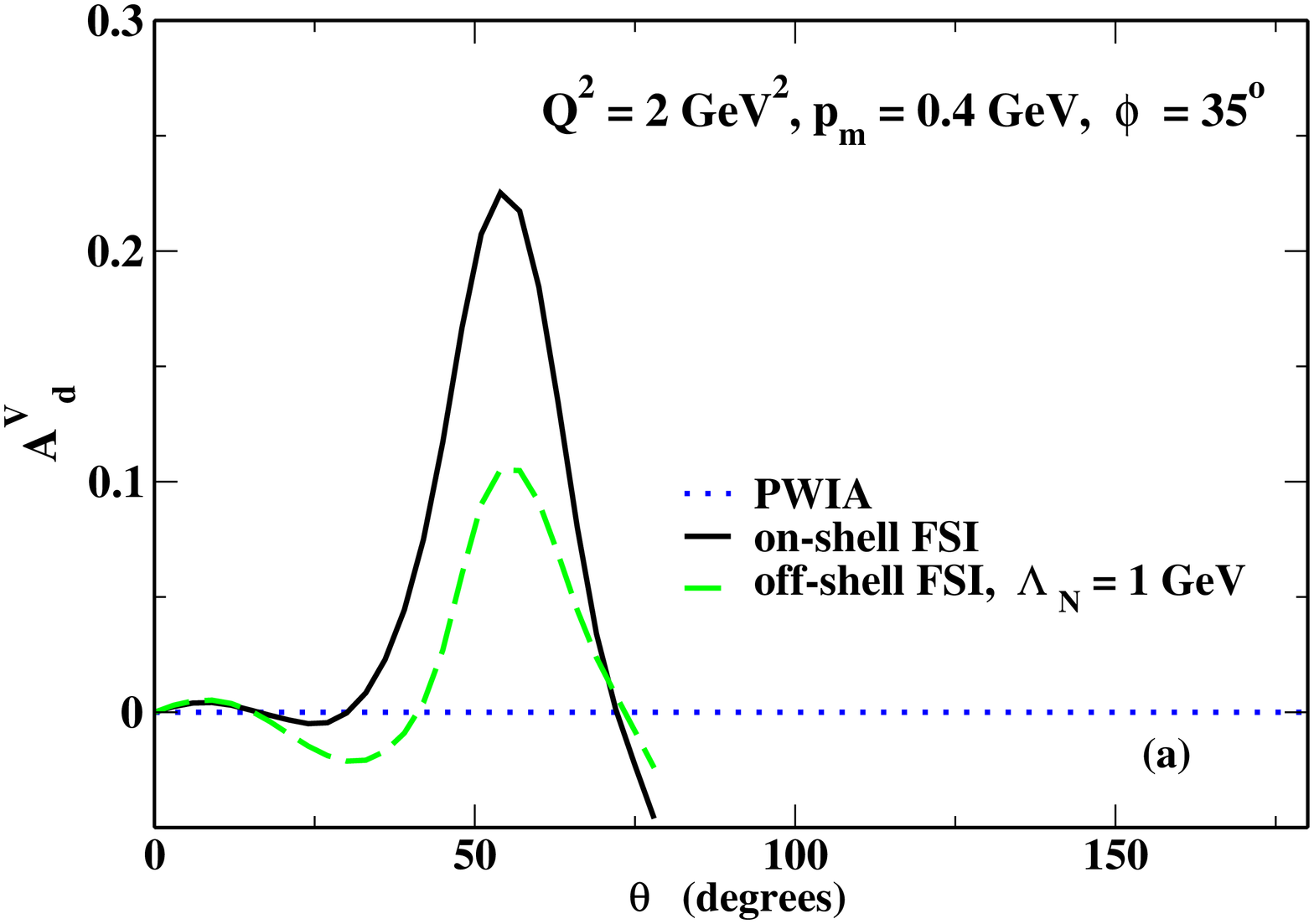}
\includegraphics[width=14pc,angle=270]{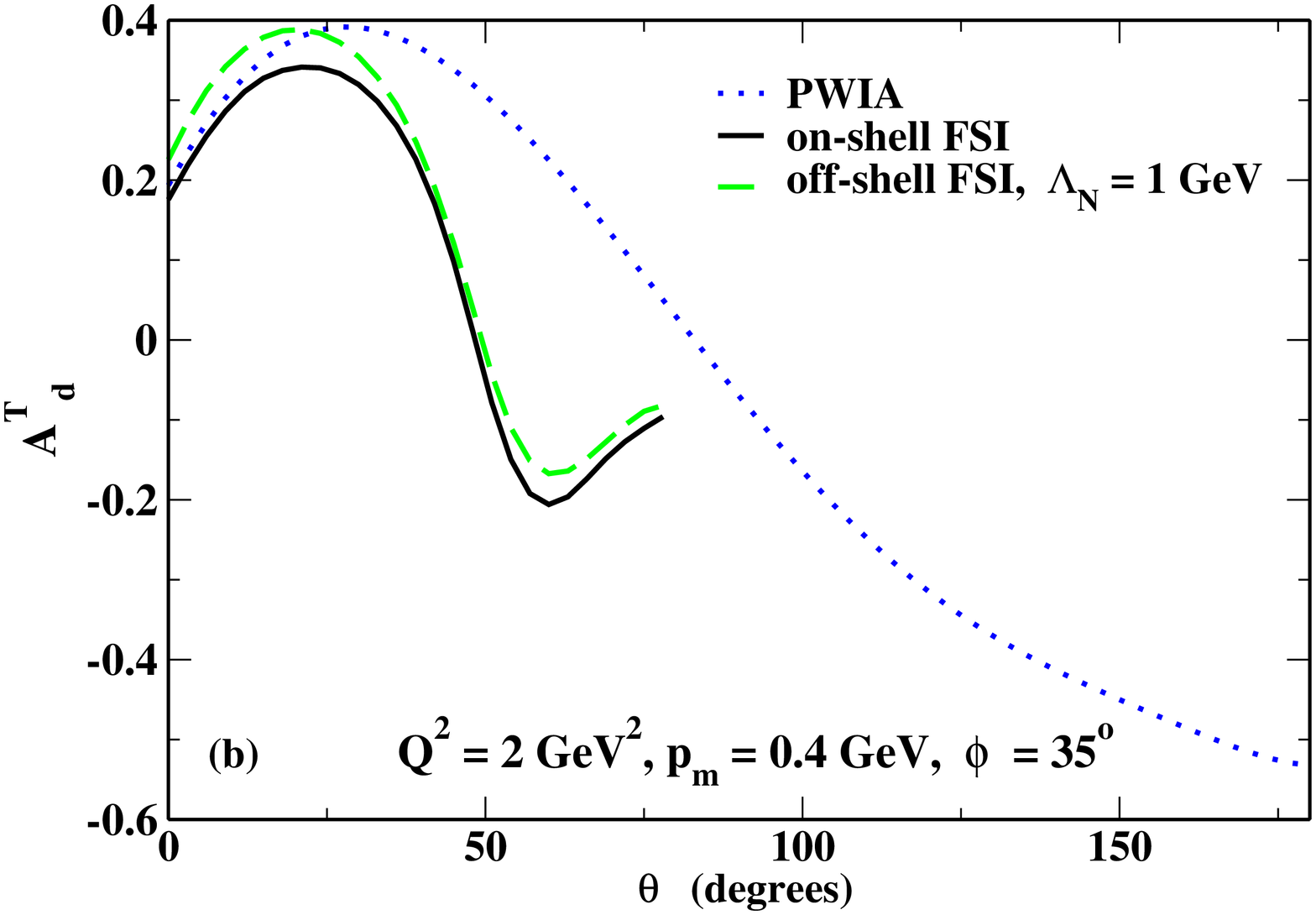}
\includegraphics[width=14pc,angle=270]{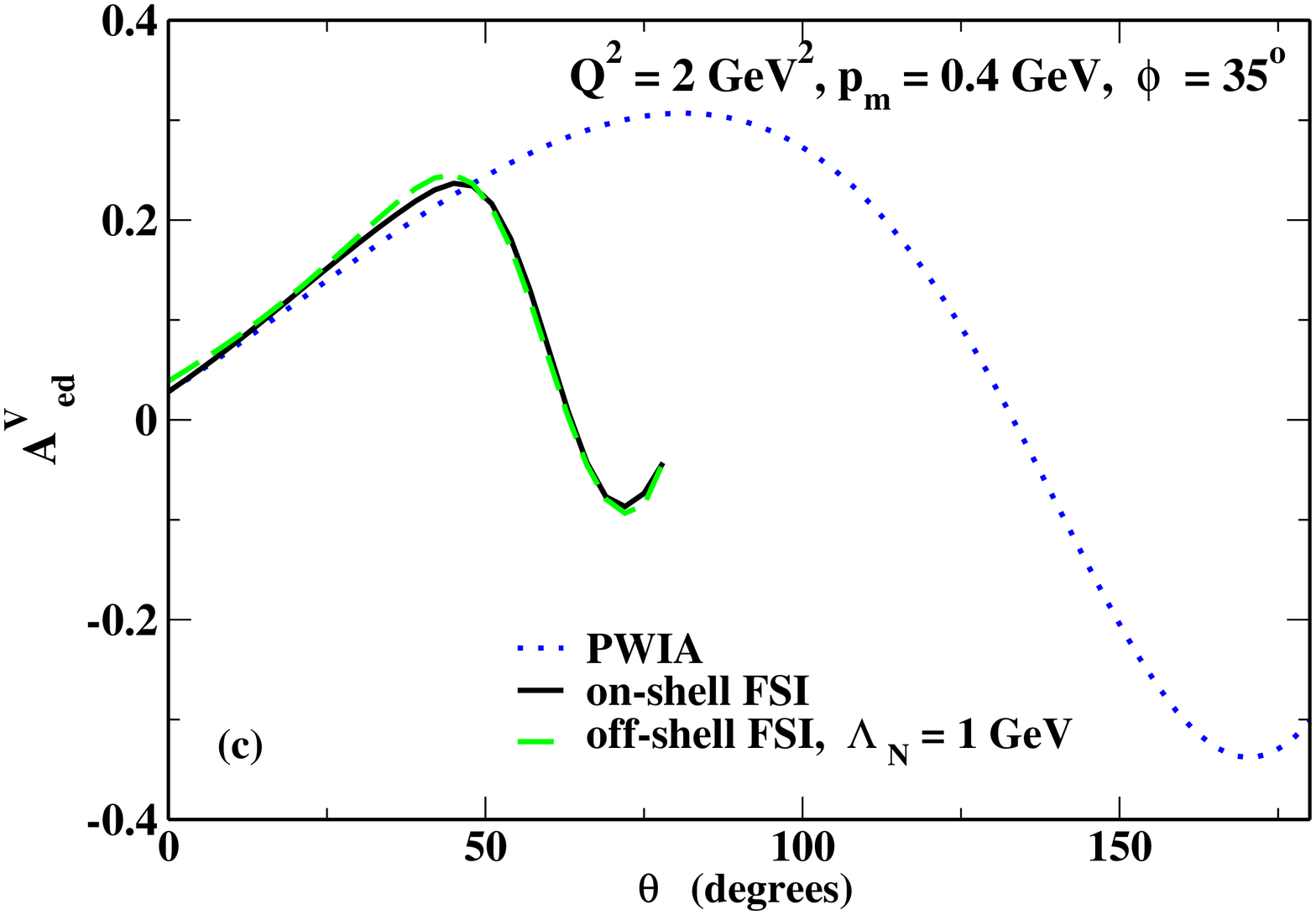}
\includegraphics[width=14pc,angle=270]{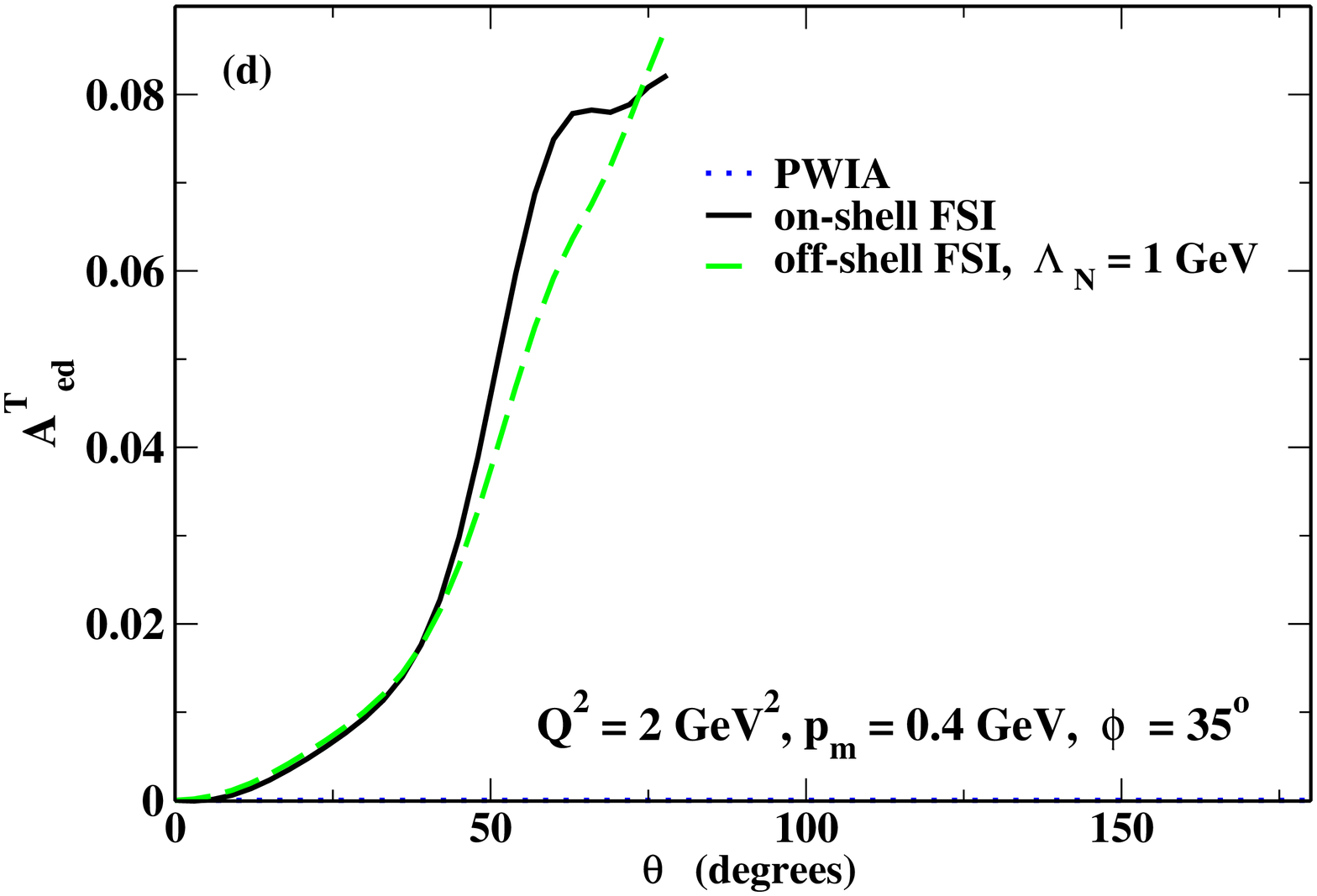}
\caption{(Color online) The asymmetries $A^V_d$ (panel (a)), $A^T_d$ (panel (b)), $A^V_{ed}$ (panel (c)) and $A^T_{ed}$ (panel (d))
for a beam energy of $5.5~
{\rm GeV}$, $Q^2 = 2 ~{\rm GeV}^2$, $p_m = 0.4$ GeV, and $\phi_p = 35^\circ$ are shown
calculated in PWIA (dotted line), with on-shell FSI (solid line), and with
on-shell and off-shell FSI (dashed line),
 as a function of the polar angle of the missing momentum. }
\label{fig_ad_pm04_pwiafsi}
\end{figure}

In Fig. \ref{fig_ad_pm04_pwiafsi}, we show the four asymmetries as functions of the angle for a fixed missing
momentum value of
$p_m = 0.4$ GeV, and for a fixed $\phi_p = 35^o$.
The non-relativistic, factorized PWIA prediction for $A^T_d$ is $A^T_d \propto 1 - 3 \cos^2 \theta$, which leads to
zeros for
$\theta = 54.7^o$ and $\theta = 125.3^o$. If we use only the S-wave and D-wave contributions to the
ground state wave function, and perform the PWIA calculation with the quantization axis
along the three-momentum transfer, $\vec q$, we observe exactly this type of angular dependence.
The P-wave contributions lead to slight deviations from the non-relativistic angular pattern.
In the figures we show, we have used a
quantization axis along the beam, and this rotation obscures the original structure of the asymmetry. 
%In practice, experimentalists polarize their targets with respect to the beam.
The angular dependence of $A^V_{ed}$ even in non-relativistic PWIA is more complicated than the structure for $A^T_d$, as the beam-vector asymmetry
$A^V_{ed}$ is equal to the ratio of helicity-dependent and helicity independent responses. This prevents the
cancelations of helicity-independent expressions in numerator and denominator that is present in the tensor asymmetry
$A^T_d$, and causes its simple angular structure.
For $A^V_d$ and $A^T_{ed}$, the non-relativistic result predicts zero for all angles, and this result persists for our
fully relativistic calculation, for the reasons discussed above.

Again, we observe the same pattern that was apparent for the momentum distributions: for $A^T_d$ and $A^V_{ed}$,
the FSI effects are small for small angles, and become important only for larger angles.
The differences between on-shell FSI calculations
and FSI calculations including off-shell FSIs, too, is very small.
For $A^V_d$ and $A^T_{ed}$, the off-shell FSI effects are more pronounced,
in particular for $A^V_{d}$. Note that when calculating an angular distribution for a fixed missing momentum,
we slice through various values of $x$, and therefore the relative importance of the off-shell FSI contributions
is different for different angles $\theta$.

At the four-momentum transfer of $2$ GeV$^2$, we are limited in the range of polar angles $\theta$ that we may access, as
complete $np$ scattering amplitudes are available only up to $1.3$ GeV. The results for $Q^2$ = 1 GeV$^2$ and
otherwise identical kinematics are not qualitatively different from
what we see at smaller angles.

\begin{figure}[ht]
\includegraphics[width=14pc,angle=270]{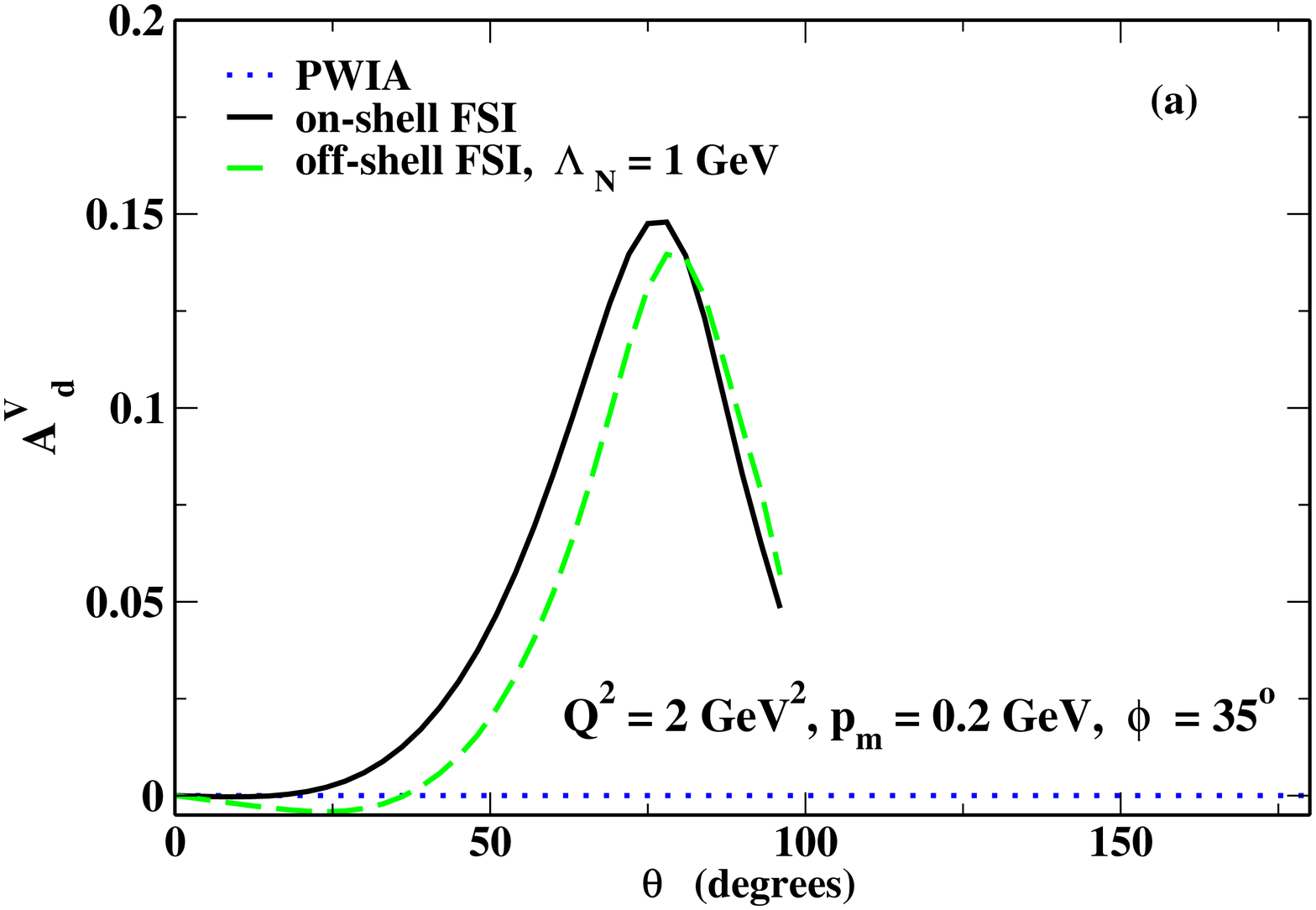}
\includegraphics[width=14pc,angle=270]{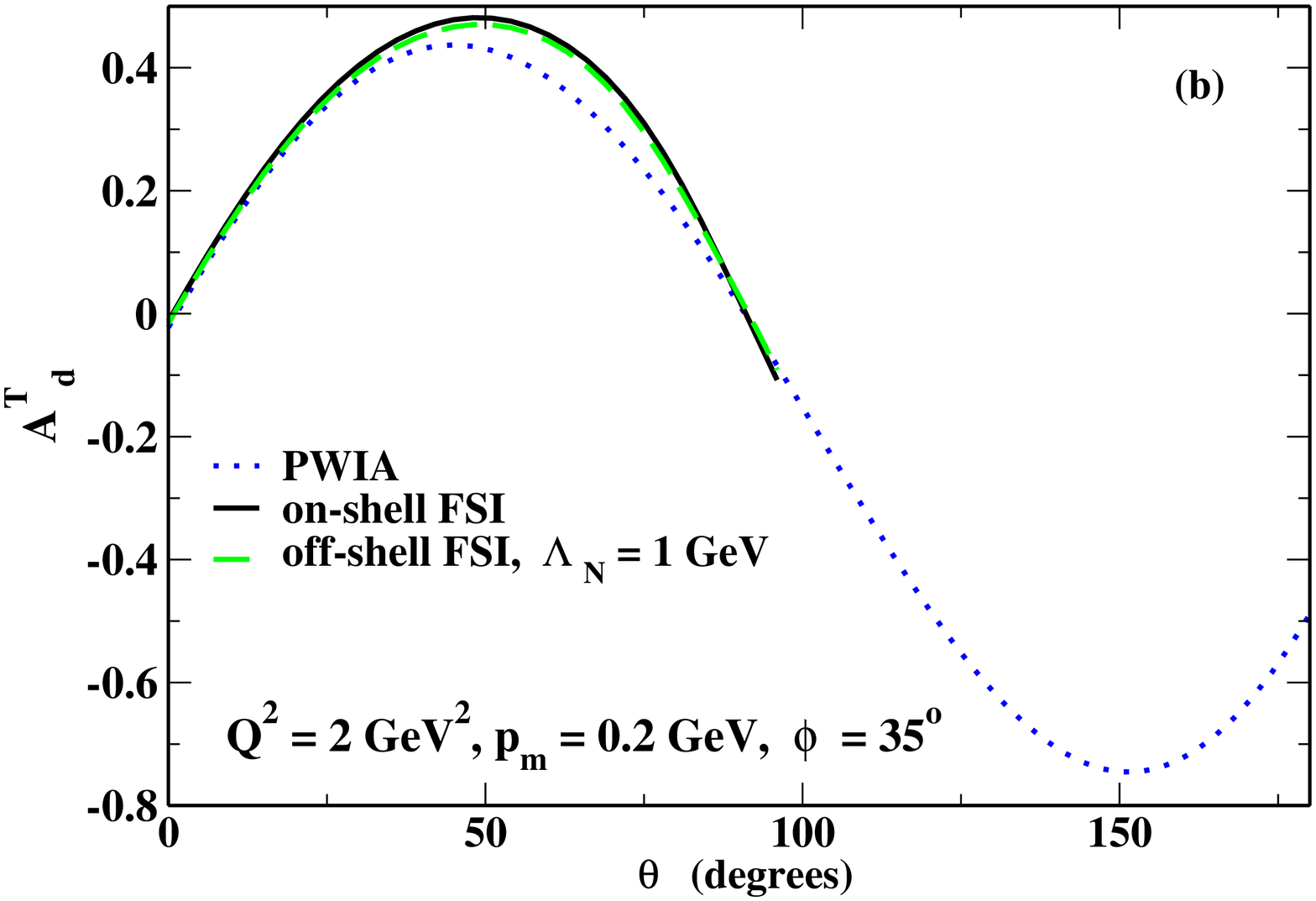}
\includegraphics[width=14pc,angle=270]{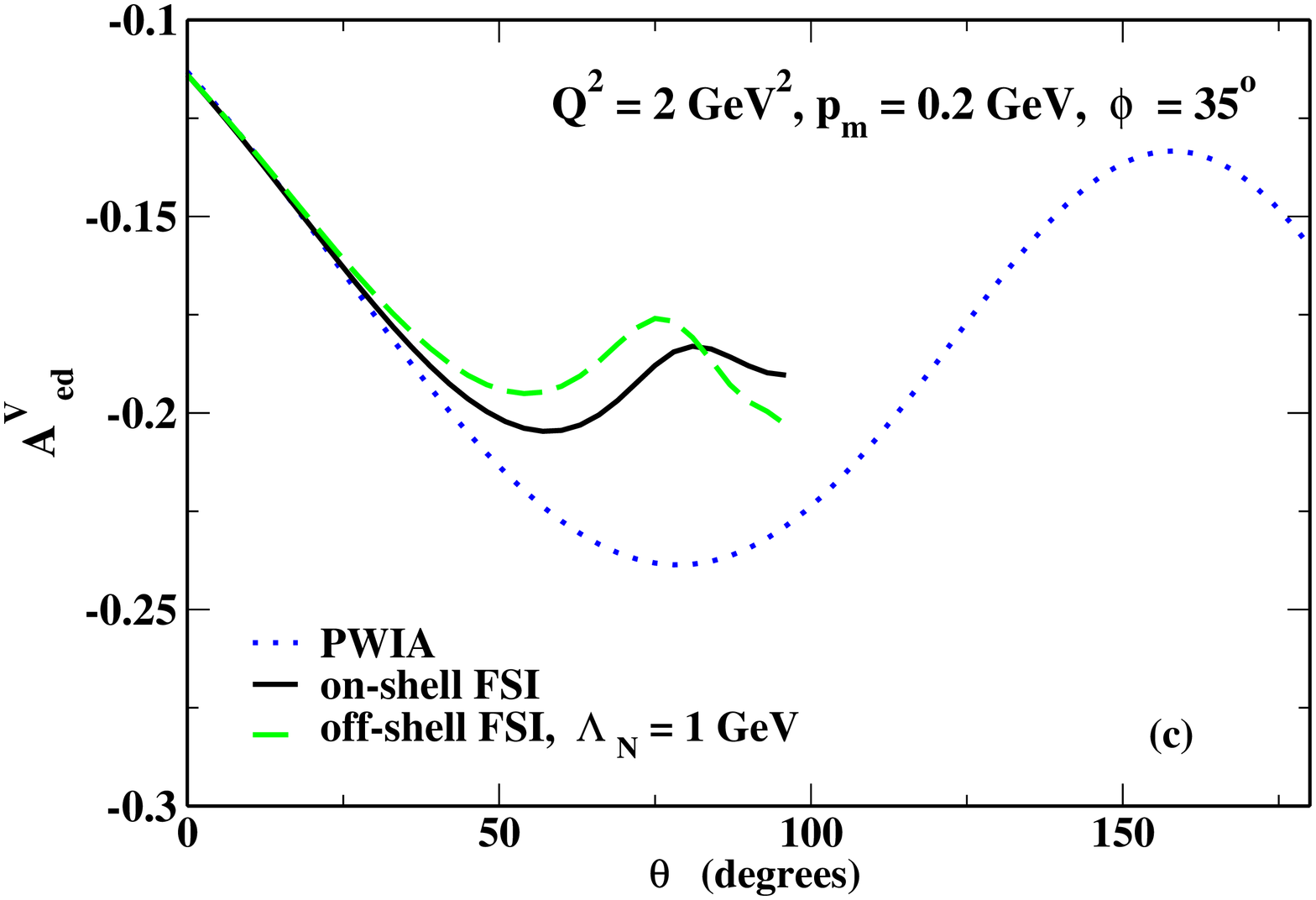}
\includegraphics[width=14pc,angle=270]{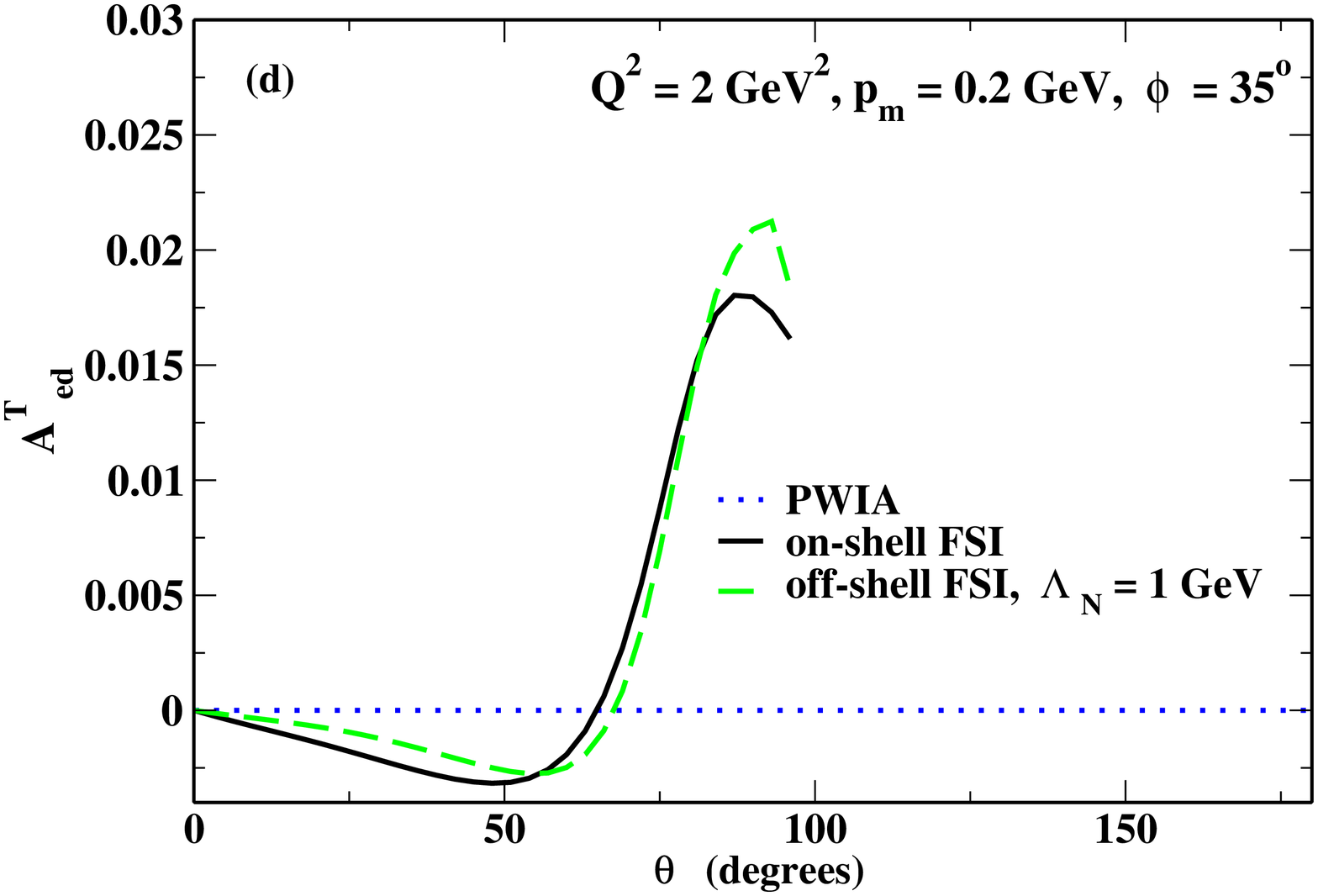}
\caption{(Color online) The asymmetries $A^V_d$ (panel (a)), $A^T_d$ (panel (b)), $A^V_{ed}$ (panel (c)) and $A^T_{ed}$ (panel (d))
for a beam energy of $5.5~
{\rm GeV}$, $Q^2 = 2 ~{\rm GeV}^2$, $p_m = 0.2$ GeV, and $\phi_p = 35^\circ$ are shown
calculated in PWIA (dotted line), with on-shell FSI (solid line), and with
on-shell and off-shell FSI (dashed line),
 as a function of the polar angle of the missing momentum. }
\label{fig_ad_pm02_pwiafsi}
\end{figure}

We show the asymmetries at a lower missing momentum value, $p_m = 0.2$ GeV, as function of the angle
in Fig.\ref{fig_ad_pm02_pwiafsi}.
Overall, it is clear that for the lower missing momentum value, $p_m = 0.2$ GeV, the influence of FSIs is not that large.
As before, the asymmetry $A^V_{d}$ shows an off-shell FSI result that differs
from the on-shell FSI for a larger range of angles, but the effect is much less pronounced than for higher $p_m$.
Both $A^V_{ed}$ and $A^T_{ed}$ show small off-shell FSI effects in the region where the asymmetries are large.
The off-shell FSI contributions are somewhat limited here, as for $p_m = 0.2$ GeV, the maximum kinematically possible
$x$ value is 1.3.

Summarizing, it is interesting to note that the tensor asymmetry $A^T_d$ and the double spin asymmetry $A^V_{ed}$ exhibit
rather similar behavior, even though they have quite different structures: the former depends on the helicity
independent terms
of the cross section and has a tensor ($T_{20}$) structure, whereas the latter depends on the helicity-dependent terms
of the cross section, and has a vector structure ($T_{10}$). Due to invariance under parity and time reversal,
both responses are non-zero in PWIA,
and they show similar structures and sensitivity
to FSI effects. Their $\phi_p$ dependence is similar, too, showing a mirror symmetry along $\phi_p = 180^o$.
In the same way, the target spin
asymmetry $A^V_d$ (helicity independent, vector) and the tensor-beam
asymmetry $A^T_{ed}$ (helicity-dependent, tensor) show similar traits: they are both zero in PWIA,
and are more sensitive even to off-shell FSI effects. Their $\phi_p$ dependence leads to an inversion
of all features above $\phi_p = 180^o$.

\subsubsection{Contributions from individual parts of the NN scattering amplitude to the FSIs}

In our calculation of the final state interactions, we use the full nucleon-nucleon scattering amplitude.
There are several ways to decompose and parametrize the
$NN$ scattering amplitude. It can be parametrized with five terms: a central, spin-independent term, a spin-orbit term, and three
double-spin flip contributions. It can also be given in terms of invariants, using a scalar, vector, tensor, pseudoscalar, and axial
term. Some of these parametrizations may be more or less useful and enlightening in trying to understand what is happening.
As we are interested in the effects of target polarization, investigating the effects of spin-dependent terms in the FSIs
is a logical and interesting step. We separate the NN amplitudes into a central term, a single spin-flip (i.e. spin-orbit) term,
and three double spin-flip terms.

\begin{figure}[ht]
\includegraphics[width=14pc,angle=270]{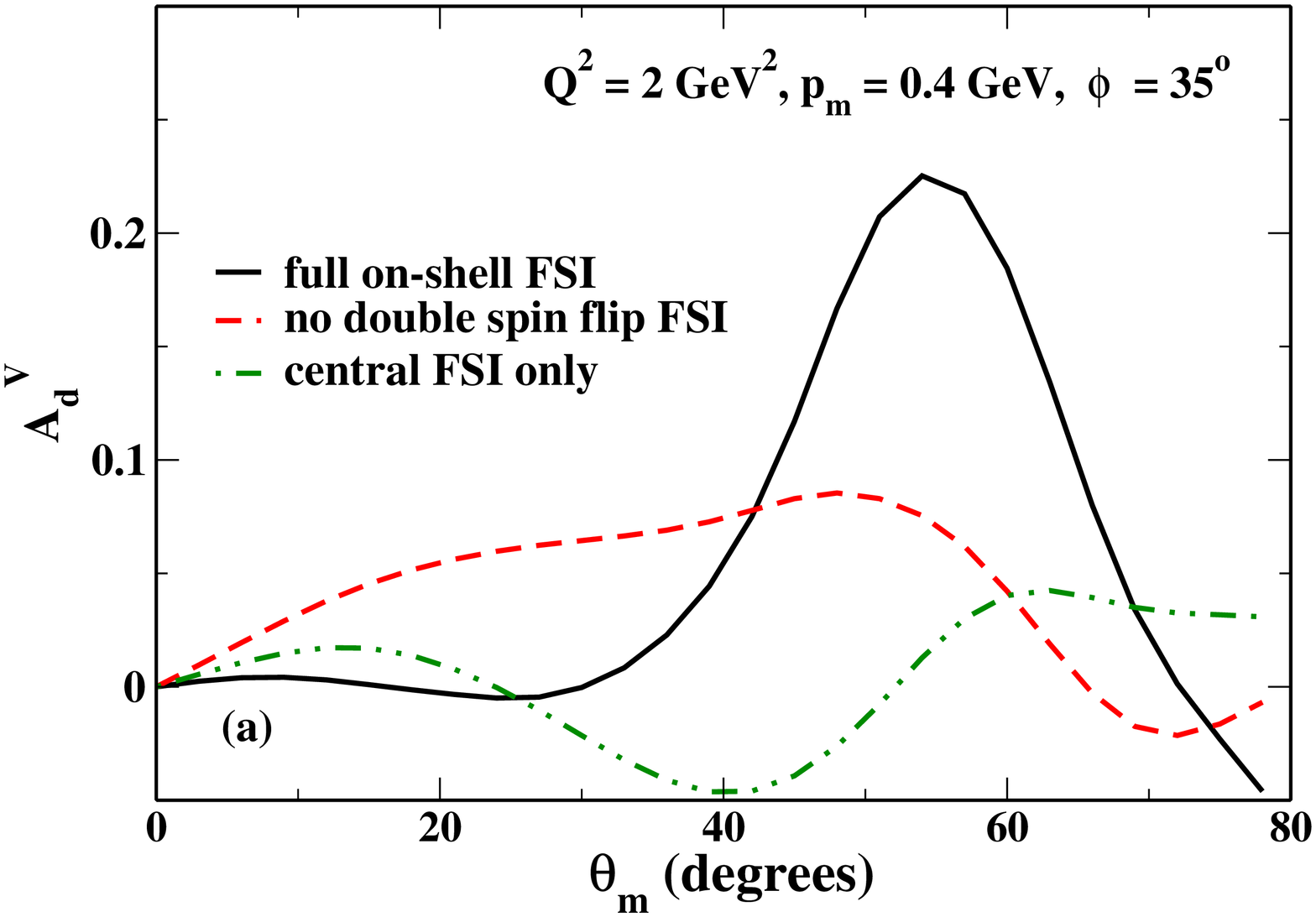}
\includegraphics[width=14pc,angle=270]{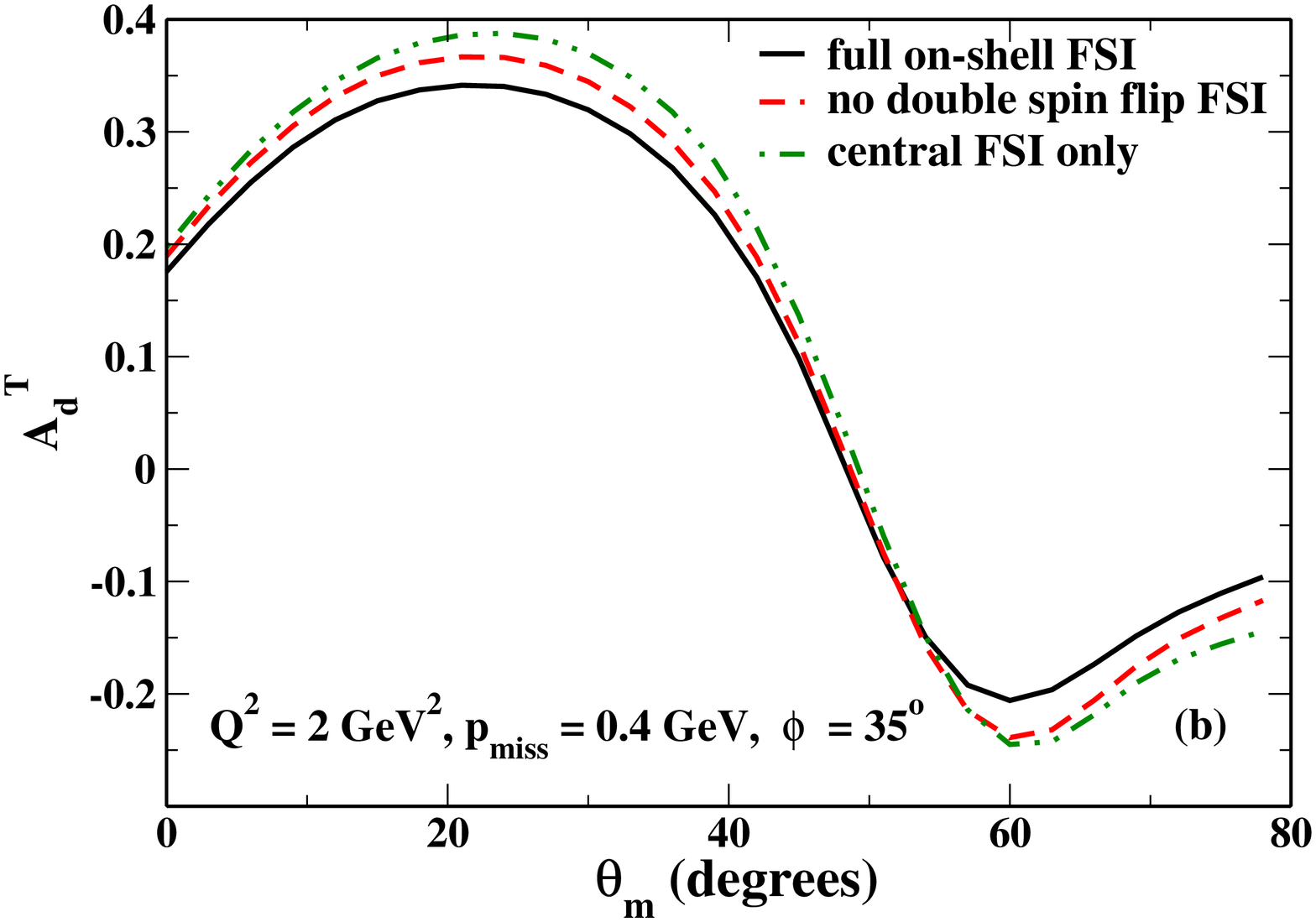}
\includegraphics[width=14pc,angle=270]{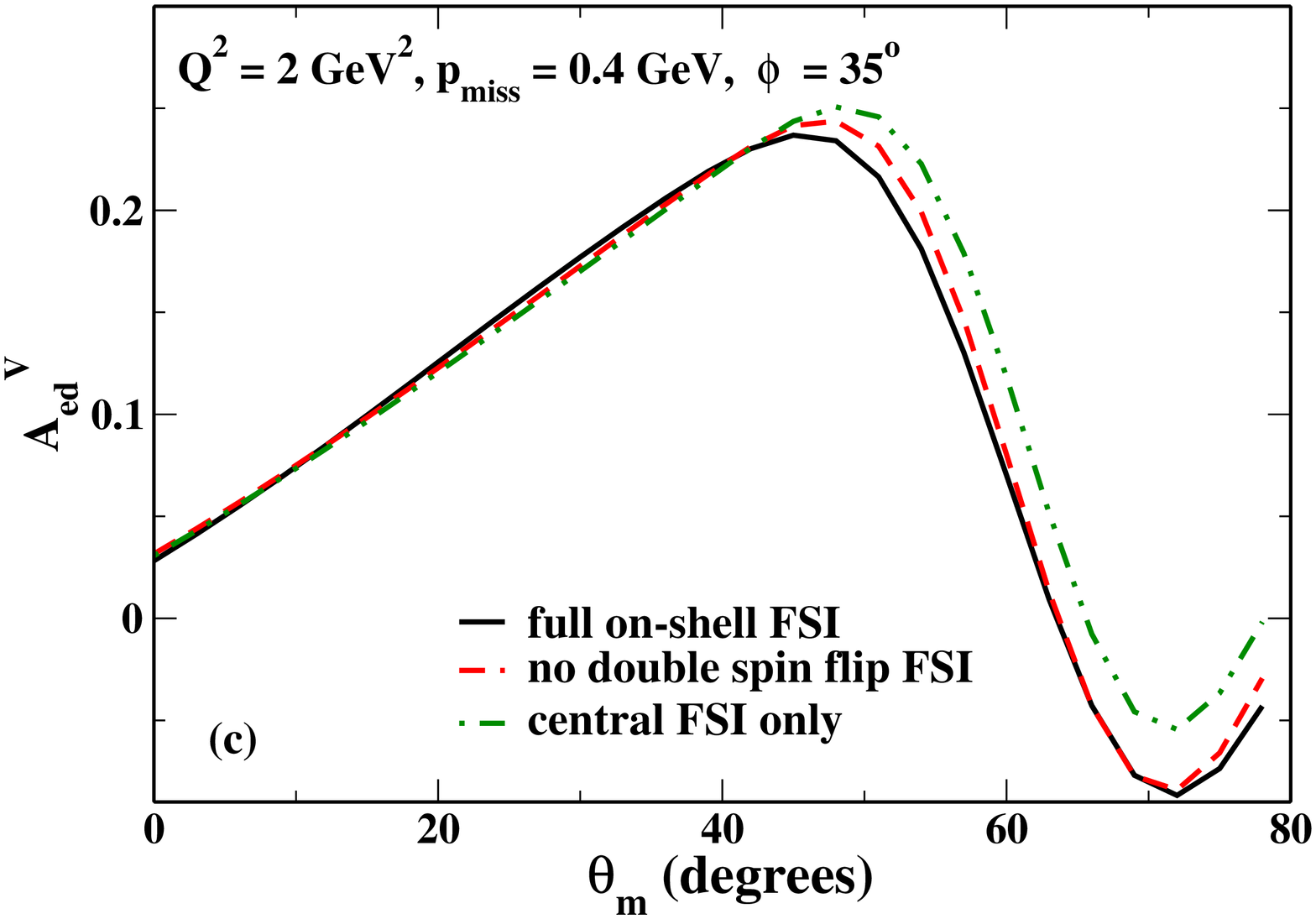}
\includegraphics[width=14pc,angle=270]{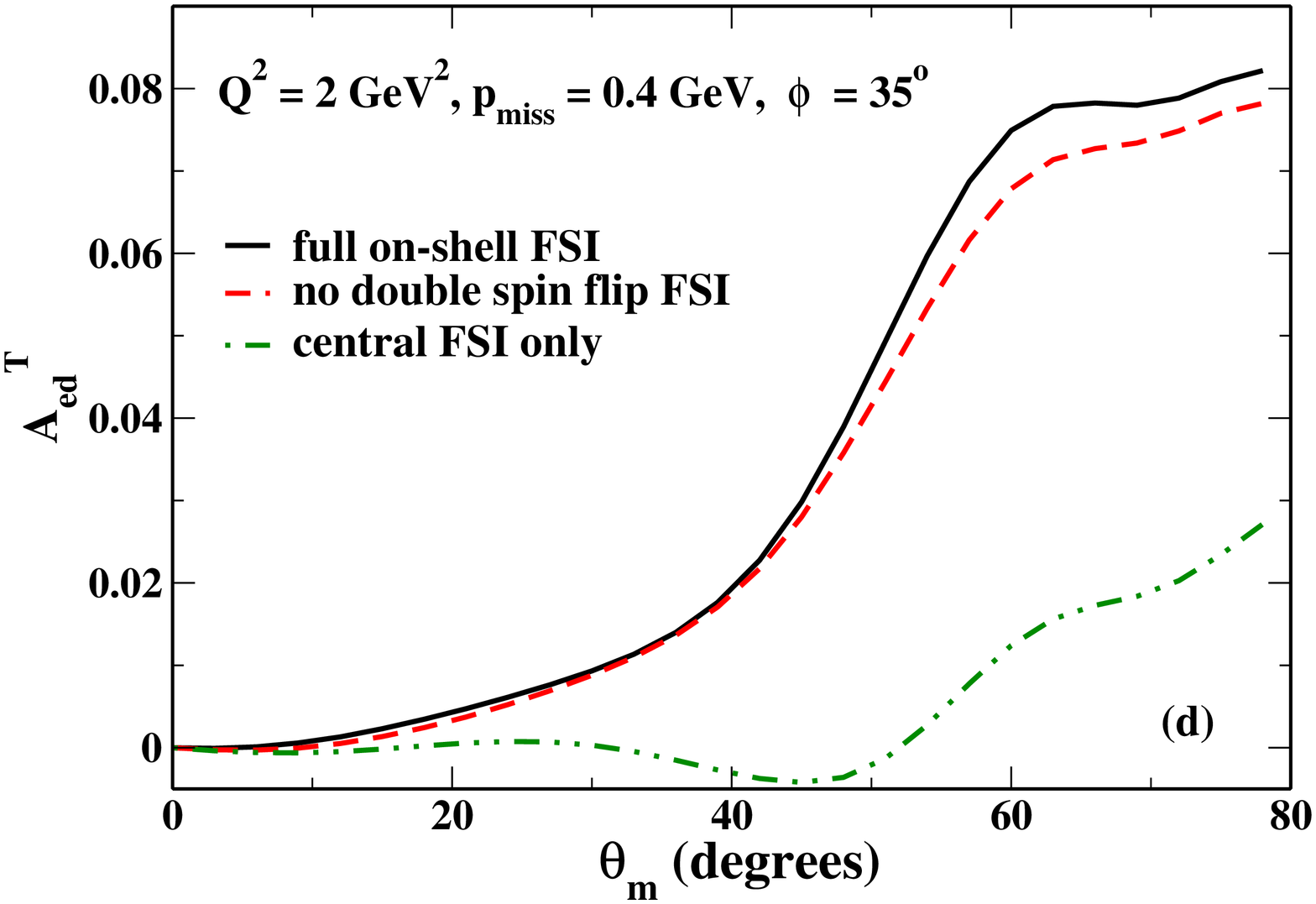}
\caption{(Color online)
The asymmetries $A^V_d$ (panel (a)), $A^T_d$ (panel (b)), $A^V_{ed}$ (panel (c)) and $A^T_{ed}$ (panel (d))
for a beam energy of $5.5~{\rm GeV}$, $Q^2 = 2 ~{\rm GeV}^2$, $p_{m} = 0.4$ GeV, and $\phi_p = 35^\circ$ are shown
calculated with on-shell FSI (solid line), without any double spin-flip
terms in the on-shell FSI (dashed line), and with central on-shell FSI only (dash-double-dotted line),
as a function of the angle $\theta_m$ of the missing momentum.}
\label{fig_ad_spindep}
\end{figure}

Fig. \ref{fig_ad_spindep} shows the contributions of the central, central and single spin-flip, and full FSIs to the
four asymmetries at $Q^2 = 2$ GeV$^2$ and $p_m = 0.4$ GeV as a function of the angle of the missing momentum.
The tensor asymmetry $A^T_d$ shows little sensitivity to the details of the FSIs, it just shows some minor quantitative
changes in the dip and peak region. The beam-vector asymmetry, $A^V_{ed}$, is insensitive at lower angles, but shows
small changes in magnitude at larger angles. In both cases, there is no shape change when the different parts
of the FSIs are added.
This changes when considering the  target-spin asymmetry $A^V_d$ and the tensor-beam asymmetry $A^T_{ed}$.
For these asymmetries,
the shape is quite different when only the central part of the FSIs is included. The central FSI result for $A^T_{ed}$
is rather small and even takes some negative values in a shallow dip around $\theta \approx 45^o$.
 Once the single spin-flip FSI is included, the asymmetry changes
and shows a steep rise with a shallow shoulder at larger angles.
 With the inclusion of the double-spin flip FSIs, the magnitude of the asymmetry increases a bit at larger angles.

For $A^V_d$, the influence of the spin-dependent FSIs is most pronounced: while the asymmetry is
very small and changes sign twice with central FSIs, the inclusion of the single spin-flip term leads to
an asymmetry that is similar in shape, albeit a bit larger than with central FSIs, and of opposite sign.
The double spin-flip terms completely change the shape of the asymmetry, leading to a pronounced peak
and a much larger maximum value. Here, for $A^V_d$, the effect of the double spin-flip FSIs is most pronounced
and most relevant.

\begin{figure}[ht]
\includegraphics[width=14pc,angle=270]{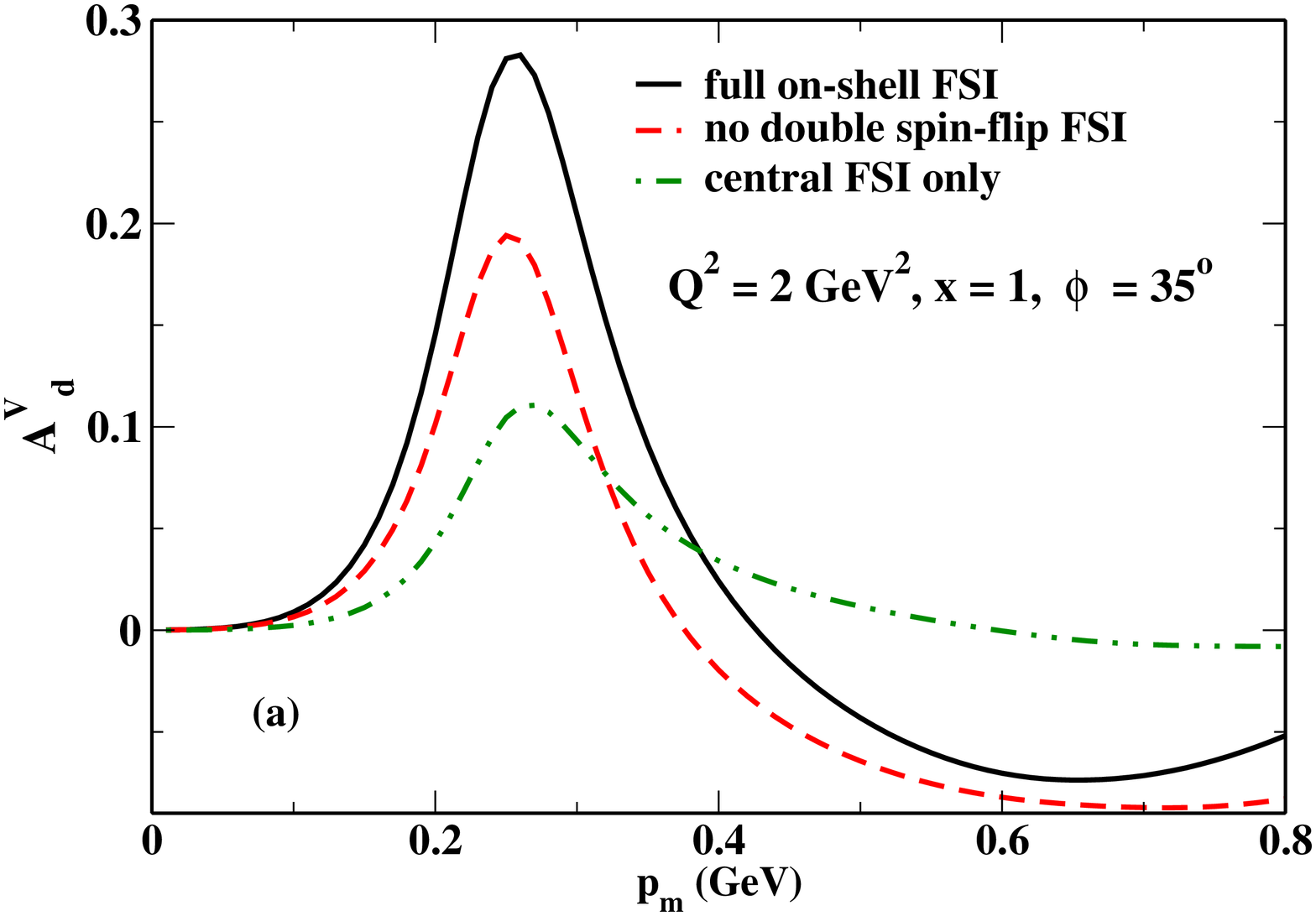}
\includegraphics[width=14pc,angle=270]{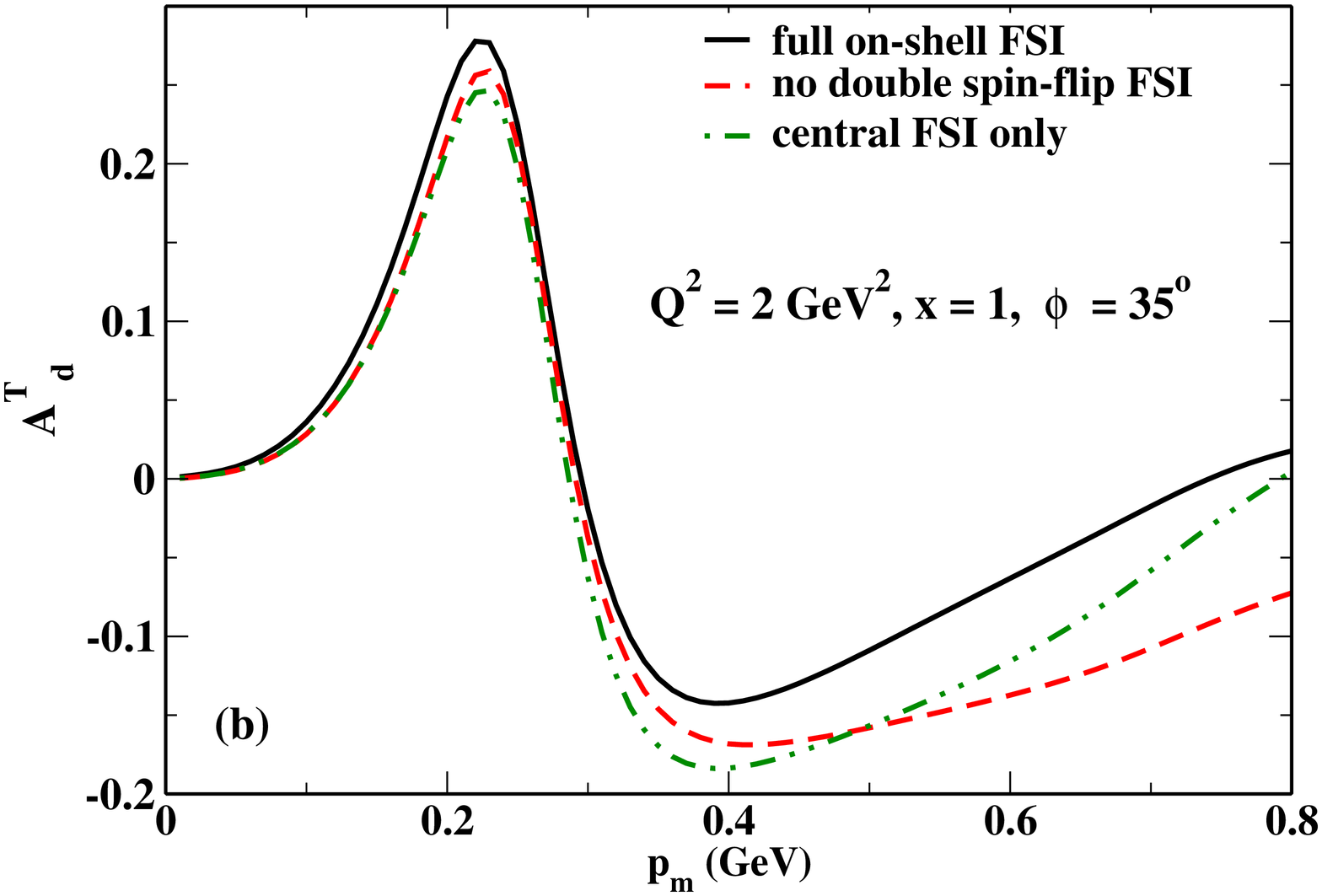}
\includegraphics[width=14pc,angle=270]{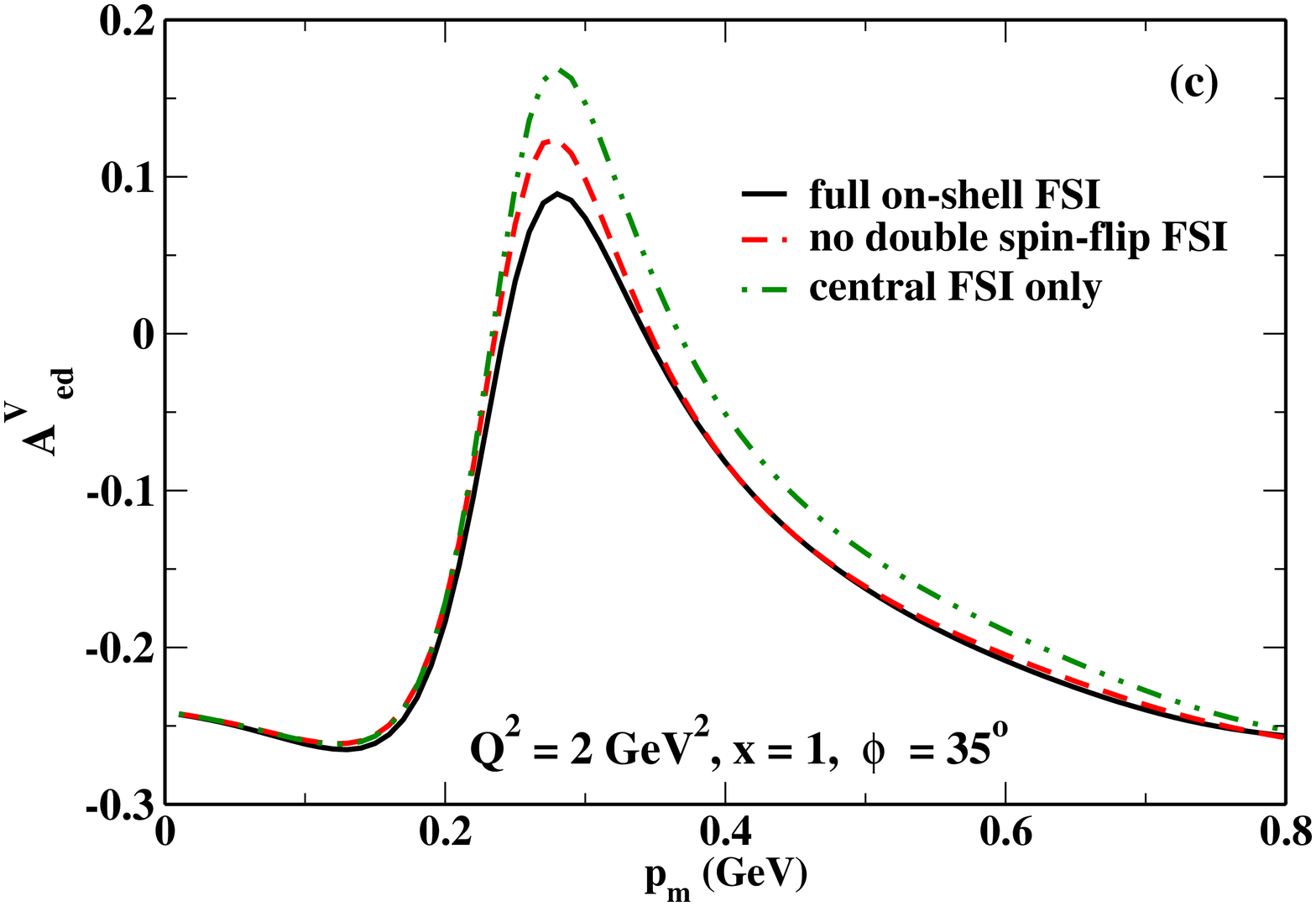}
\includegraphics[width=14pc,angle=270]{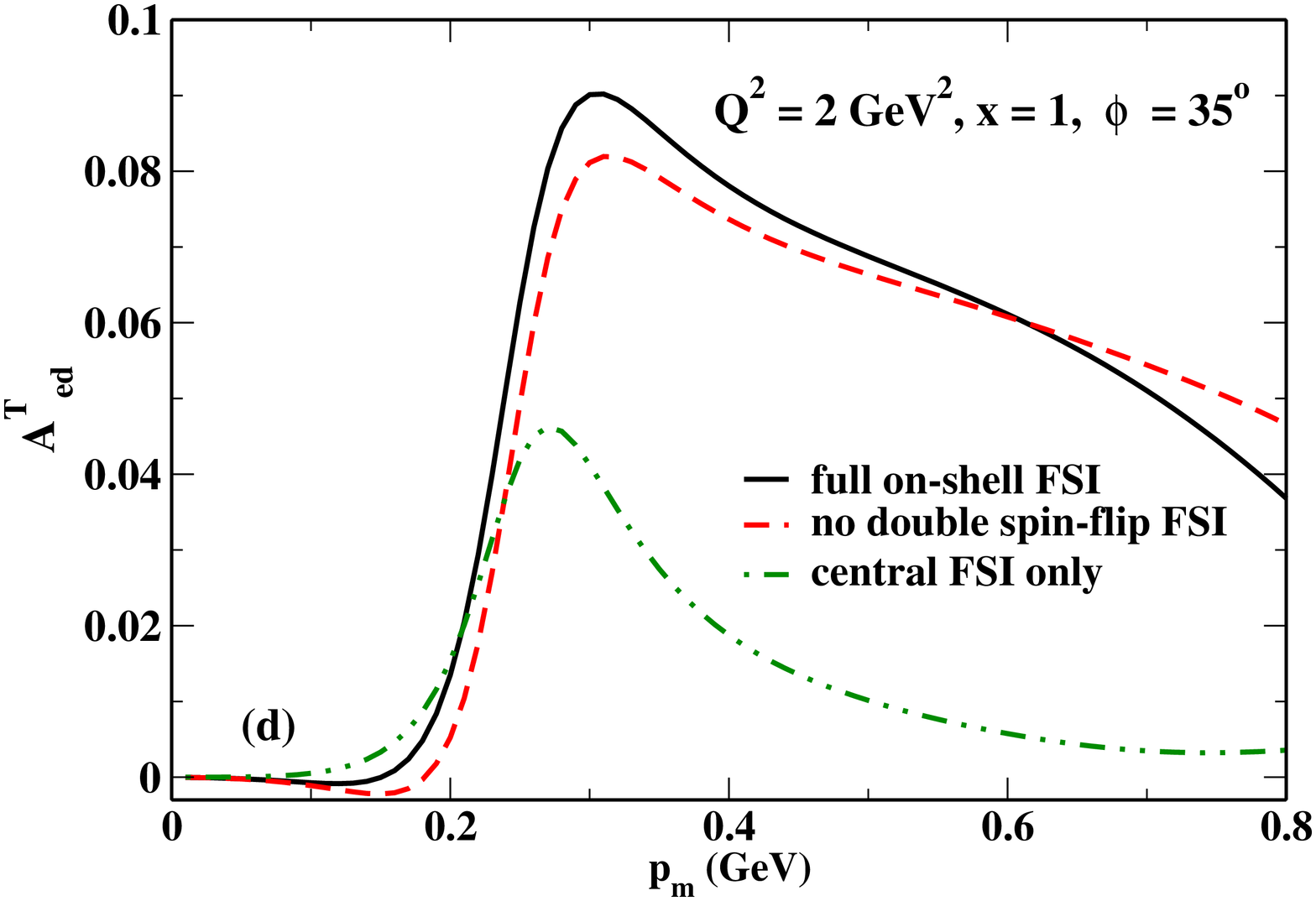}
\caption{(Color online)
The asymmetries $A^V_d$ (panel (a)), $A^T_d$ (panel (b)), $A^V_{ed}$ (panel (c)) and $A^T_{ed}$ (panel (d))
for a beam energy of $5.5~{\rm GeV}$, $Q^2 = 2 ~{\rm GeV}^2$, $x = 1$, and $\phi_p = 35^\circ$ are shown
calculated with the full on-shell FSI (solid line), without any double spin-flip
terms in the on-shell FSI (dashed line), and with central on-shell FSI only (dash-double-dotted line)
 as a function of the missing momentum.}
\label{fig_md_spindep}
\end{figure}

Fig. \ref{fig_md_spindep} shows the contributions of the central, central and single spin-flip, and full FSIs to the
four asymmetries at $Q^2 = 2 $GeV$^2$ and $x = 1$ as a function of the missing momentum. While the angular distributions
for $p_m = 0.4$ GeV shown in Fig. \ref{fig_ad_spindep} do not show a very pronounced effect of the double
spin-flip terms on $A^T_d$ and $A^V_{ed}$, the momentum distribution for $A^T_d$ shows that especially for larger
missing momenta, the spin-dependent FSIs are very relevant. Starting for $p_m = 0.4$ GeV, the results without the full
spin-dependence deviate significantly from the full FSI result, and for $p_m > 0.5$ GeV, an interesting inversion
happens: the result for central FSI only is closer to - but still far off - the full FSI result than the calculation
without double spin-flip. This indicates that interference effects are relevant for $A^T_d$ in this kinematic region.
For $A^V_d$, a similar picture emerges for larger missing momentum: for $p_m > 0.4$ GeV, central FSI only results
are above the full result, while the no double spin-flip result is below it. For this asymmetry, for $p_m > 0.1$ GeV,
all types of spin-dependent FSI are very important. The tensor-beam asymmetry, $A^T_{ed}$, shows that while the
double spin-flip terms have only a small effect, the single spin-flip term, i.e. the spin-orbit term, gives
a huge contribution.

Overall, the tensor asymmetry $A^T_d$ and
the double spin asymmetry $A^V_{ed}$ exhibit rather similar behavior, showing some quantitative and no large
qualitative dependence on spin-dependent FSIs. The target-spin asymmetry $A^V_d$ and the tensor-beam asymmetry $A^T_{ed}$
show large qualitative and quantitative sensitivity to spin-dependent FSIs, each in a different way.

In a previous paper \cite{bigdpaper} dealing with unpolarized observables, we investigated the influence of
the different invariant amplitudes of
the $NN$ amplitude parametrization by calculating the FSIs with only one of the invariant amplitudes. For the unpolarized
case, we found that the role of interference is huge, and that there is no single dominant amplitude.
For the asymmetries, we find that for small angles, the pseudoscalar
amplitude seems to be very close to all asymmetries except for $A^T_{ed}$, but this behavior is confined to
$\theta < 30^o$. Deviations beyond that are significant, in particular for $A^V_d$. The results show that overall,
there are many relevant
interference effects, and no single part of the $NN$ amplitude is dominant.

\subsubsection{Influence of the D wave}

A question often discussed is the influence of correlations in the nuclear ground state, and in the
case of the deuteron, the role played by D-wave - and P-wave - admixtures. Due to the rather different
predictions of various non-relativistic $NN$ models for the D-wave content, the hunt for observables sensitive
to this part of the wave function has been going on for a long time. It should be noted that on theoretical grounds the attempts to extract the D-wave contribution to the deuteron bound state is ill considered.  The wave function is not an observable and unitary transformations can change the D-state contribution while leaving the matrix elements unchanged. Thus an actual observable contains information about initial and final states, as well as on the
current operator, with the various quantities changing with unitary transformation and one cannot be uniquely separated from the others.

In our calculation, due to the normalization of the ground state wave function, there are
some issues with directly isolating the D-wave contribution. Just in order to give an impression
of the influence of the D-wave contribution on the asymmetries we study here, we have simply switched off
the D-wave contributions, without changing the normalizations. In our relativistic calculation, there is also
a P-wave contribution present. We study its effect, too. In general, P-wave effects are expected to be very small.

\begin{figure}[ht]
\includegraphics[width=19pc,angle=0]{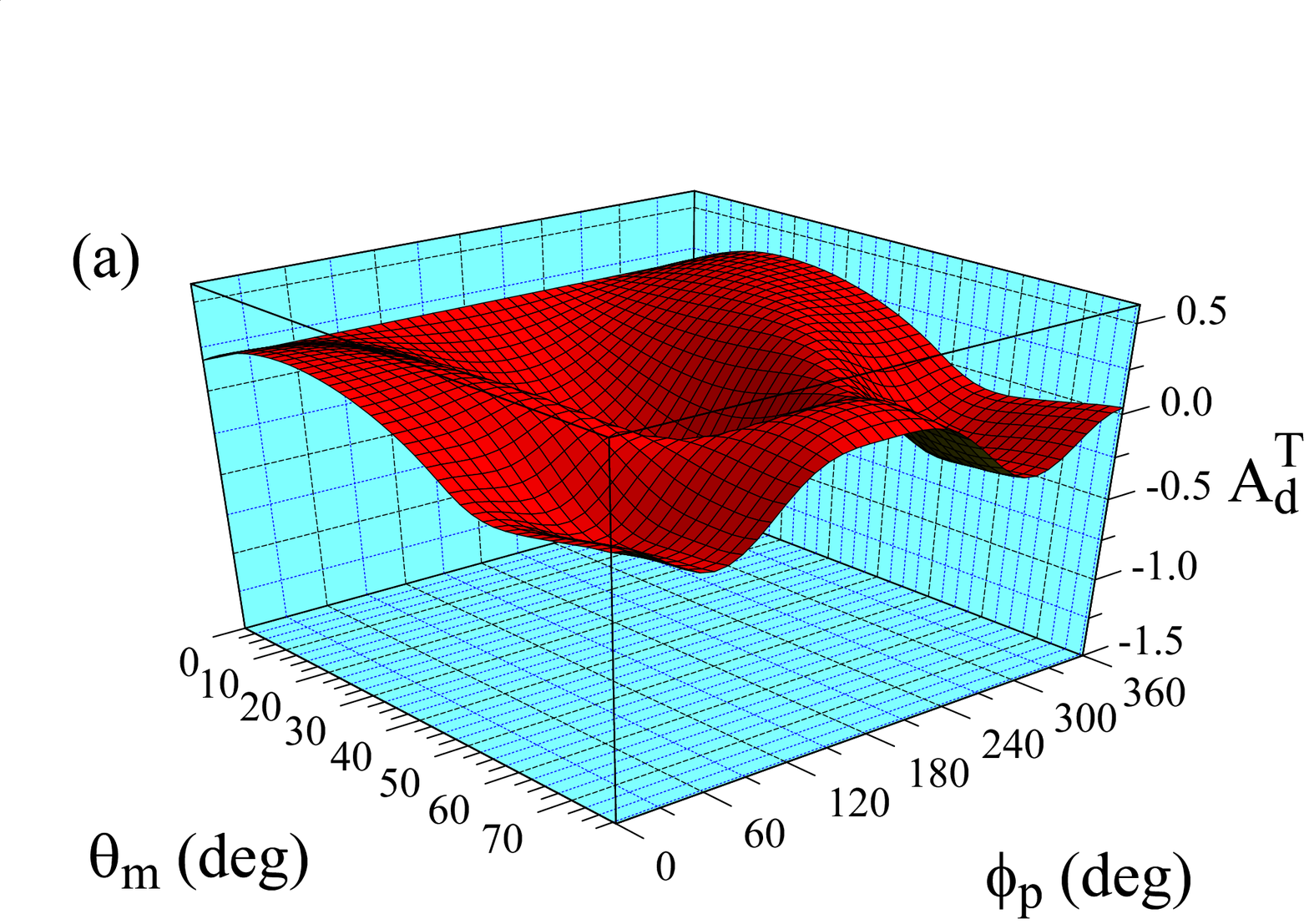}
\includegraphics[width=19pc,angle=0]{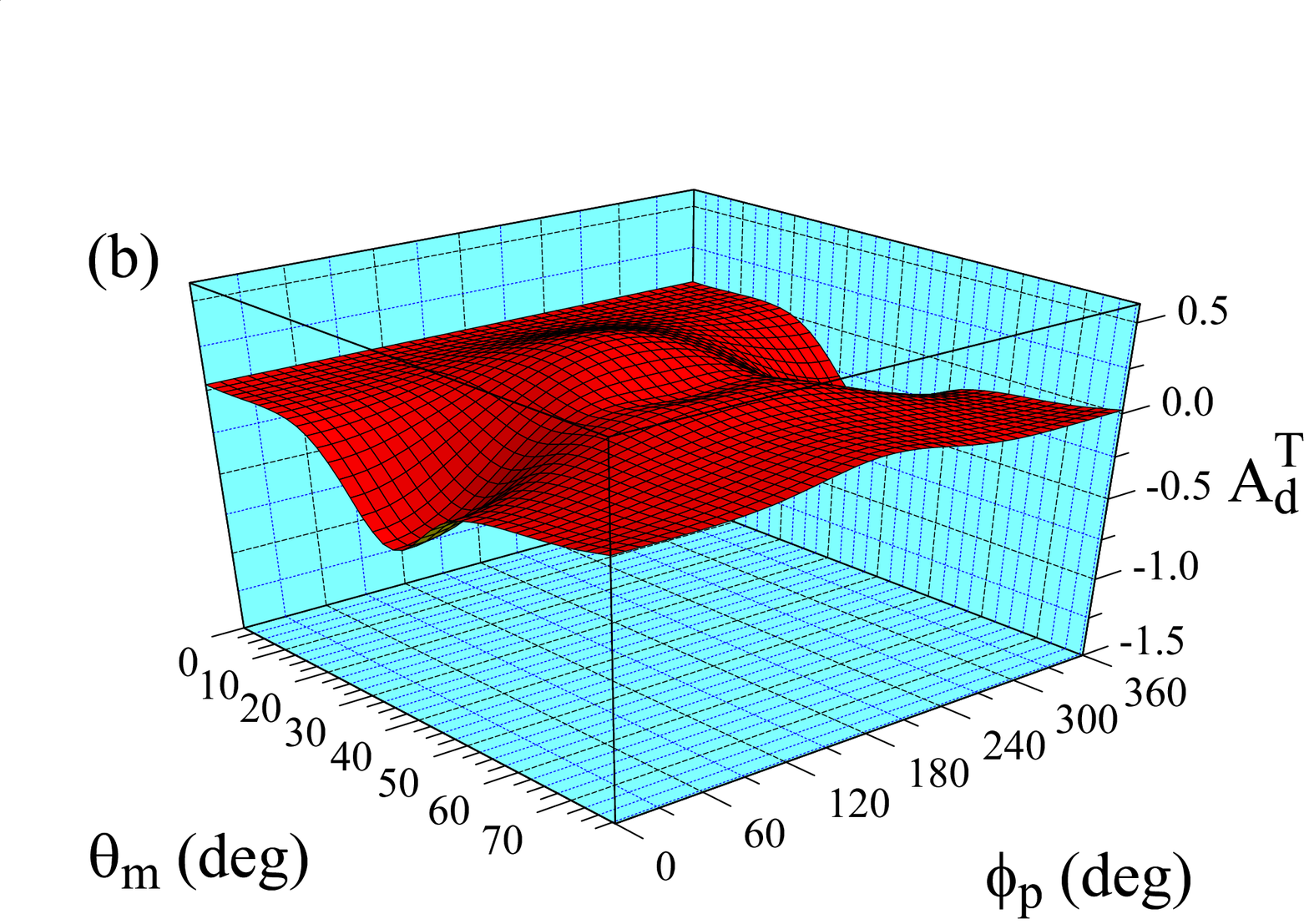}
\includegraphics[width=19pc,angle=0]{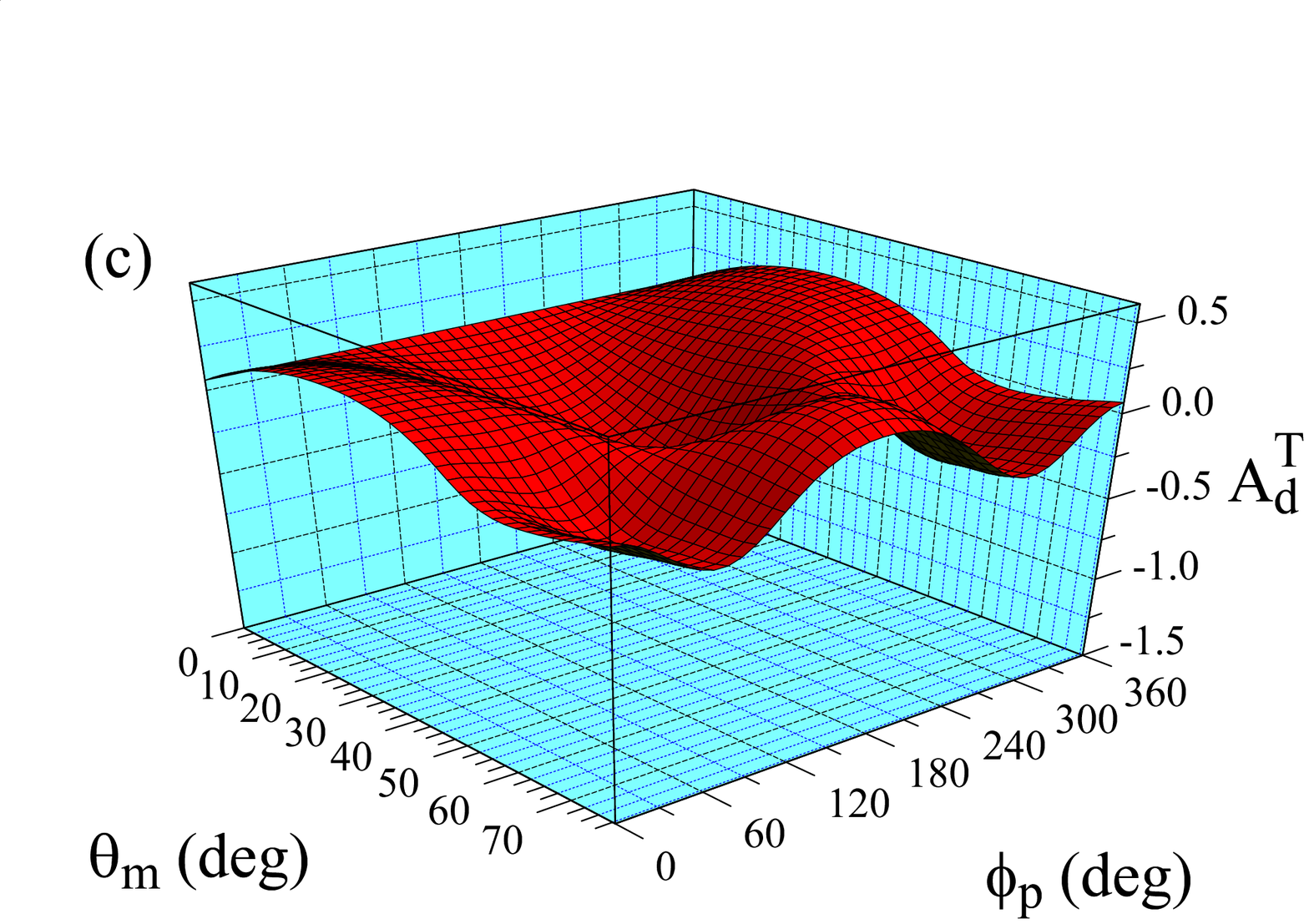}
\includegraphics[width=19pc,angle=0]{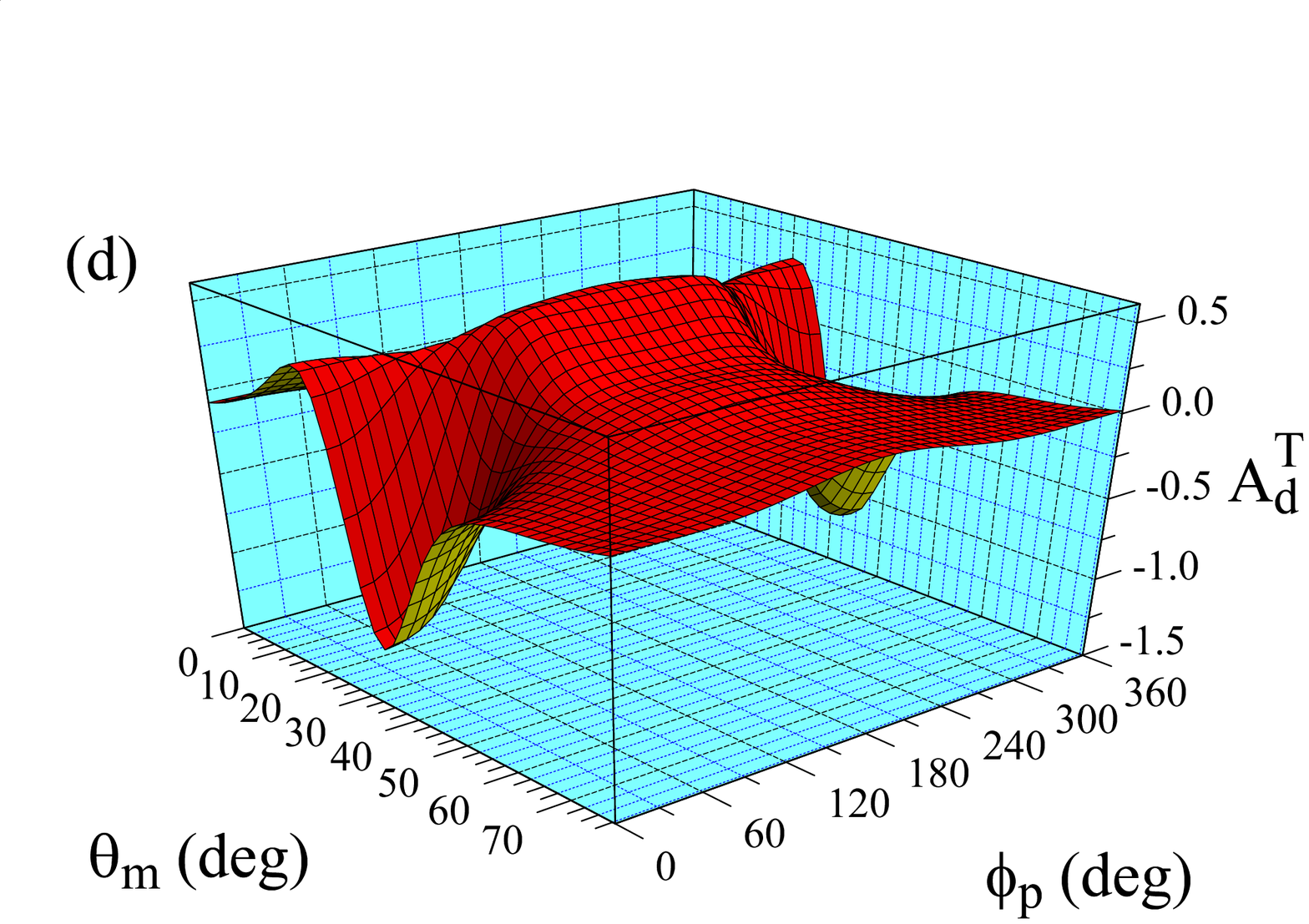}
\caption{(Color online) The asymmetry $A^T_d$,
for a beam energy of $5.5~
{\rm GeV}$, $Q^2 = 2 ~{\rm GeV}^2$, and $p_m = 0.4~{\rm GeV}$ is shown
calculated (a) with on-shell FSI, (b) without a D-wave contribution to the ground state,
(c) without a P-wave contribution to the ground state,
and (d) with only an S-wave contribution to the ground state,
 as a function of polar angle of the missing momentum $\theta_m$ and the proton's azimuthal angle. }
\label{fig_3d_multi_atd_qsq2pm04_spdwave_fsi}
\end{figure}

While the other asymmetries also show significant dependence on the D-wave,
we will focus here for brevity on the effects of the D-wave and P-wave contributions on the tensor asymmetry $A^T_d$.
As expected, the difference between the calculation with the full ground-state wave function (a)
and the calculation without the D-wave (b) is large: a prominent dip is turned
into a peak, and the maximum values reached change. In non-relativistic PWIA, this asymmetry
would be zero without the D-wave, but with FSI - even just central FSI - the tensor asymmetry acquires a
nonzero value, as the relative position of the neutron and the knocked-out proton matter for the strength of the
FSI that is experienced.

Performing a calculation without the P-wave contribution (c) does not lead to any significant changes, the peak heights
vary a little, but there are no qualitative changes. Panel (d) shows the results for just
the S-wave part of the wave function.  Here, the
missing P-wave contribution - still present in the top right panel without the D-wave - leads to a somewhat different
shape and an increased magnitude for the dip structure at lower $\phi_p$ values.
It is interesting to note that in PWIA, if we switch off the D-wave contribution, the tensor asymmetry is
small but still non-zero due to the P-wave contributions.

\section{Summary and Outlook}

In this paper, we have presented a formalism for the calculation of responses and asymmetries for polarized
deuteron targets. We have shown how to evaluate these observables in different reference frames, and for
different polarization axes. Symmetries of the current matrix elements were pointed out,
and together with the behavior under parity and time reversal transformations, exploited to show
that two of the asymmetries we discuss,  the target-spin asymmetry $A^V_d$ and
the tensor-beam asymmetry $A^T_{ed}$, vanish in PWIA.

We performed a relativistic calculation of various asymmetries accessible with a polarized
deuteron target. We have included a full FSI calculation, with on-shell and off-shell
contributions, using experimental data on the $pn$ scattering amplitude as input. Final state interactions
are very relevant for all asymmetries in most kinematics.
Two of the asymmetries vanish in PWIA, and are therefore more sensitive to FSI effects, even to the off-shell
FSI contributions. An important result
of our paper is that even in the region of the quasi-elastic ridge, $x = 1$, the influence of FSIs on the
asymmetries is large, and a straightforward extraction of D-wave properties from measured data will not be
possible. This is true even though the influence of the D-wave on the asymmetries is large, as commonly assumed.
The influence of the P-waves, a purely relativistic phenomenon, is generally small, unless we consider situations where
the D-wave is switched off.

One interesting and conspicuous feature of the asymmetries is the fact that the target-spin asymmetry $A^V_d$ and
the tensor-beam asymmetry $A^T_{ed}$ have very similar properties  - they vanish in PWIA, have
similar sensitivity to FSIs, and a similar
dependence on $\phi_p$ - just as the tensor asymmetry $A^T_d$ and the beam-vector asymmetry $A^V_{ed}$ show
similar properties in these respects. We have shown that this can be understood in terms of their behavior under
parity and time reversal in PWIA.

We have tested the sensitivity of our results to the different parts of the FSIs. As expected, spin-dependent FSIs are
relevant, and depending on the kinematics and observable, even the double spin-flip terms are extremely important.

Our calculation has been performed in impulse approximation, i.e. assuming that the detected proton
is the nucleon that interacted with the photon initially. Contributions from the Born term, where the
photon interacts with the neutron, will in general be small for most observables in most kinematics,
but they may contribute noticeably for larger missing momenta ($p_m > 0.6$ GeV).

Next, we plan to perform calculations for an unpolarized deuteron target and a polarized ejected nucleon.

{\bf Acknowledgments}:  We thank Sebastian Kuhn for his insightful comments on an
earlier version of this paper. We thank Michael Kohl for providing us with information on
the Bates experiments. This work was
supported in part by funds provided by the U.S. Department of Energy
(DOE) under cooperative research agreement under No.
DE-AC05-84ER40150 and by the National Science Foundation under
grant No. PHY-0653312.

\appendix
\section{The Density matrix}

Consider an object with total angular momentum $j$ and  projection $m$. An arbitrary state of angular momentum $j$ can be written as
\begin{equation}
\left|j\right>_i=\sum_{m=-j}^j c^i_m\left|jm\right>
\end{equation}
where normalization of the state requires that
\begin{equation}
\sum_{m=-j}^j \left|c^i_m\right|^2=1\,.
\end{equation}
The expectation value of some operator $\hat{A}$ for this state is given by
\begin{equation}
<\hat{A}>_i=\sum_{m'=j}^j{c^i_{m'}}^*\left<jm'\right|\hat{A}\sum_{m=-j}^j\left|jm\right>c^i_{m}
=\sum_{m'=j}^j\sum_{m=-j}^j\left<j,m'\right|\hat{A}\left|jm\right>c^i_{m}{c^i_{m'}}^*\,.
\end{equation}

Any attempt to polarize a target consisting of a collection of these objects by applying magnetic fields will in general not produce a single state such as that described above, but will consist of a statistical ensemble of such states with probabilities $P_i$ such that
\begin{equation}
\sum_iP_i=1\,.
\end{equation}
The statistical average of the expectation value of operator $\hat{A}$ is then given by
\begin{equation}
<\hat{A}>=\sum_iP_i<\hat{A}>_i
=\sum_{m'=-j}^j\sum_{m=-j}^j\left<j,m'\right|\hat{A}\left|jm\right>\sum_ic^i_{m}P_i{c^i_{m'}}^*\,.
\end{equation}
Defining the density matrix
\begin{equation}
\rho_{mm'}=\sum_ic^i_{m}P_i{c^i_{m'}}^*\label{rhodef}
\end{equation}
and
\begin{equation}
A_{m'm}=\left<j,m'\right|\hat{A}\left|jm\right>
\end{equation}
the average expectation value of $\hat{A}$ can be written as
\begin{equation}
<\hat{A}>=
\sum_{m'=-j}^j\sum_{m=-j}^jA_{m'm}\rho_{mm'}={\rm Tr}(\bm{A}\bm{\rho})
\end{equation}
where $\bm{A}$ and $\bm{\rho}$ are matrix representation of $\hat{A}$ and the density matrix in the subspace of total angular momentum $j$.

From  (\ref{rhodef}),
\begin{equation}
\rho_{mm'}^*=\sum_ic^i_{m'}P_i{c^i_{m}}^*=\rho_{m'm}
\end{equation}
or in matrix form
\begin{equation}
\bm{\rho}^\dag=\bm{\rho}\,.
\end{equation}
So the density matrix is hermitian. Also,
\begin{equation}
{\rm Tr}(\bm{\rho})=\sum_{m=-j}^j\sum_ic^i_{m}P_i{c^i_{m}}^*
=\sum_i\sum_{m=-j}^j\left|c^i_{m}\right|^2P_i=\sum_iP_i=1\,.
\end{equation}
A further constraint on density matrix is given by
\begin{equation}
{\rm Tr}(\bm{\rho}^2)\leq({\rm Tr}(\bm{\rho}))^2=1\,.\label{rhoconstraint}
\end{equation}

It is often convenient to express the density matrix for angular momentum $j$ in terms of spherical tensor operators such that
\begin{equation}
\hat{\rho}=\frac{1}{2j+1}\sum_{J=0}^{2j}\sum_{M=-J}^JT^*_{JM}\hat{\tau}_{JM}
\end{equation}
where $\hat{\tau}_{JM}$ is an irreducible spherical tensor operator of rank $J$ and projection $M$ and the $T_{JM}$ are complex coefficients that describe the average polarization of the target. These are defined such that
\begin{equation}
T^*_{JM}=(-1)^MT_{J-M}\label{Tsymmetry}
\end{equation}
and
\begin{equation}
\hat{\tau}^\dag_{JM}=(-1)^M\hat{\tau}_{J-M}\,.\label{tausymmetry}
\end{equation}

Since we are concerned with a polarized deuteron target in this paper we will now confine the argument to the case $j=1$.  In this case matrix elements of the density operator are given by
\begin{equation}
\rho_{\lambda\lambda'}=\left<1\lambda\right|\hat{\rho}\left|1\lambda'\right>
=\frac{1}{3}\sum_{J=0}^2\sum_{M=-J}^JT^*_{JM}\left<1\lambda\right|\hat{\tau}_{JM}\left|1\lambda'\right>\,.
\end{equation}
This can be written in matrix form as
\begin{equation}
\bm{\rho}=\frac{1}{3}\sum_{J=0}^2\sum_{M=-J}^JT^*_{JM}\bm{\tau}_{JM}
\end{equation}
If we choose normalizations such that
\begin{equation}
T_{00}=1
\end{equation}
and
\begin{equation}
\left<1\right.\left|\right|\hat{\tau}_J\left|\right|\left. 1\right>=\sqrt{3}\sqrt{2J+1}\,,
\end{equation}
the matrices $\bm{\tau}_{JM}$ are
\begin{eqnarray}
\bm{\tau}_{00}=\left(
\begin{array}{ccc}
1 & 0 & 0\\
0 & 1 & 0\\
0 & 0 & 1
\end{array}
\right)&& \nonumber\\
\bm{\tau}_{10}=\sqrt{\frac{3}{2}}\left(
\begin{array}{ccc}
1 & 0 & 0\\
0 & 0 & 0\\
0 & 0 & -1
\end{array}
\right)&
\bm{\tau}_{11}=\sqrt{\frac{3}{2}}\left(
\begin{array}{ccc}
0 & -1 & 0\\
0 & 0 & -1\\
0 & 0 & 0
\end{array}
\right)&\nonumber\\
\bm{\tau}_{20}=\frac{1}{\sqrt{2}}\left(
\begin{array}{ccc}
1 & 0 & 0\\
0 & -2 & 0\\
0 & 0 & 1
\end{array}
\right)&
\bm{\tau}_{21}=\sqrt{\frac{3}{2}}\left(
\begin{array}{ccc}
0 & -1 & 0\\
0 & 0 & 1\\
0 & 0 & 0
\end{array}
\right)&
\bm{\tau}_{22}=\sqrt{3}\left(
\begin{array}{ccc}
0 & 0 & 1\\
0 & 0 & 0\\
0 & 0 & 0
\end{array}
\right)\label{taumatrices}
\end{eqnarray}
and the remaining matrices can be obtained from
\begin{equation}
\bm{\tau}^\dag_{JM}=(-1)^M\bm{\tau}_{J-M}\,.
\end{equation}
These matrices have the properties
\begin{equation}
{\rm Tr} (\bm{\tau}_{JM})=0
\end{equation}
and
\begin{equation}
{\rm Tr} (\bm{\tau}^\dag_{J'M'}\bm{\tau}_{JM}) = 3\,\delta_{J'J}\,\delta_{M'M}\,.
\end{equation}
So,
\begin{equation}
{\rm Tr}(\bm{\tau}^\dag_{JM}\bm{\rho})=T^*_{JM}\,.
\end{equation}
The constraint given in (\ref{rhoconstraint}) requires that
\begin{equation}
\frac{1}{3}\left(1+\sum_{J=1}^2\sum_{M=-J}^J\left|T_{JM}\right|^2\right)\leq 1\,.
\end{equation}

Using the matrices defined by (\ref{taumatrices}), we can write
\begin{equation}
\bm{\rho}^D=\frac{1}{3}\left\{\bm{1}+\sum_{J=1}^2\left[T_{J0}\bm{\tau}_{J0}
+\sum_{M=1}^J\left(T^*_{JM}\bm{\tau}_{JM}+T^*_{J-M}\bm{\tau}_{J-M}\right)\right]\right\}\,.
\end{equation}
The last term of this can be rewritten using (\ref{Tsymmetry}) to give
\begin{eqnarray}
\sum_{M=1}^J\left(T^*_{JM}\bm{\tau}_{JM}+T^*_{J-M}\bm{\tau}_{J-M}\right)
&=&\sum_{M=1}^J\left(T^*_{JM}\bm{\tau}_{JM}+(-1)^MT_{JM}\bm{\tau}_{J-M}\right)\nonumber\\
&=&\sum_{M=1}^J\left[\Re(T_{JM})\left(\bm{\tau}_{JM}+(-1)^M\bm{\tau}_{J-M}\right)\right.\nonumber\\
&&\quad\left.+\Im(T_{JM})(-i)\left(\bm{\tau}_{JM}-(-1)^M\bm{\tau}_{J-M}\right)\right]\nonumber\\
&=&\sum_{M=1}^J\left(\Re(T_{JM})\bm{\tau}^\Re_{JM}+\Im(T_{JM})\bm{\tau}^\Im_{JM}\right)
\end{eqnarray}
where
\begin{equation}
\bm{\tau}^\Re_{JM}=\bm{\tau}_{JM}+(-1)^M\bm{\tau}_{J-M}=\bm{\tau}_{JM}+\bm{\tau}_{JM}^\dag\label{taureal}
\end{equation}
and
\begin{equation}
\bm{\tau}^\Im_{JM}=-i\left(\bm{\tau}_{JM}-(-1)^M\bm{\tau}_{J-M}\right)=-i\left(\bm{\tau}_{JM}-\bm{\tau}_{JM}^\dag\right)\,.\label{tauimaginary}
\end{equation}
The orthogonality relations for the $\bm{\tau}_{JM}$ can be used to show that
\begin{equation}
{\rm Tr}\left[{\bm{\tau}^\Re_{JM}}^\dag \bm{\rho}^D\right]=2\Re(T_{JM})
\end{equation}
and
\begin{equation}
{\rm Tr}\left[{\bm{\tau}^\Im_{JM}}^\dag \bm{\rho}^D\right]=2\Im(T_{JM})\,.
\end{equation}


\begin{thebibliography}{299}


\bibitem{cteep}
J.~Ryckebusch, W.~Cosyn, B.~Van Overmeire and C.~Martinez,
  %``The Transparency Of Nuclei To Nucleons And Pions In A Relativistic Glauber
  %Approximation,''
  Eur.\ Phys.\ J.\  A {\bf 31}, 585 (2007);
  %%CITATION = EPHJA,A31,585;%%
J.~Ryckebusch, P.~Lava, M.~C.~Martinez, J.~M.~Udias and
J.~A.~Caballero,
  %``Nuclear Transparencies In Relativistic A(E, E-Prime P) Models,''
  Nucl.\ Phys.\  A {\bf 755}, 511 (2005);
  P.~Lava, M.~C.~Martinez, J.~Ryckebusch, J.~A.~Caballero and J.~M.~Udias,
  %``Nuclear transparencies in relativistic A(e,e' p) models,''
  Phys.\ Lett.\  B {\bf 595}, 177 (2004)
  [arXiv:nucl-th/0401041];
L.~L.~Frankfurt, W.~R.~Greenberg, G.~A.~Miller, M.~M.~Sargsian and M.~I.~Strikman,
  %``Color transparency effects in electron deuteron interactions at
  %intermediate Q**2,''
  Z.\ Phys.\  A {\bf 352}, 97 (1995)
  [arXiv:nucl-th/9501009].
  %%CITATION = ZEPYA,A352,97;%%

\bibitem{hallbgmn}
J.~Lachniet {\it et al.}  [CLAS Collaboration],
  %``A Precise Measurement of the Neutron Magnetic Form Factor GMn in the
  %Few-GeV2 Region,''
  Phys.\ Rev.\ Lett.\  {\bf 102}, 192001 (2009)
  [arXiv:0811.1716 [nucl-ex]].

\bibitem{misakneutstar}
  L.~Frankfurt, M.~Sargsian and M.~Strikman,
  %``Recent observation of short range nucleon correlations in nuclei and their
  %implications for the structure of nuclei and neutron stars,''
  Int.\ J.\ Mod.\ Phys.\  A {\bf 23}, 2991 (2008)
  [arXiv:0806.4412 [nucl-th]].

\bibitem{wallyreview}
M.~Garcon and J.~W.~Van Orden,
  %``The deuteron: Structure and form factors,''
  Adv.\ Nucl.\ Phys.\  {\bf 26}, 293 (2001)
  [arXiv:nucl-th/0102049].
  %%CITATION = ANUPB,26,293;%%


\bibitem{ronfranz}
  R.~A.~Gilman and F.~Gross,
  %``Electromagnetic structure of the deuteron,''
  J.\ Phys.\ G {\bf 28}, R37 (2002)
  [arXiv:nucl-th/0111015].
  %%CITATION = JPHGB,G28,R37;%%

\bibitem{sickreview}
  I.~Sick,
  %``Elastic Electron Scattering from Light Nuclei,''
  Prog.\ Part.\ Nucl.\ Phys.\  {\bf 47}, 245 (2001)
  [arXiv:nucl-ex/0208009].
  %%CITATION = PPNPD,47,245;%%

\bibitem{bigdpaper}
S.~Jeschonnek and J.~W.~Van Orden,
%``A new calculation for D(e,e'p)n at GeV energies,''
Phys.\ Rev.\  C {\bf 78}, 014007 (2008).


\bibitem{wallyfranzwf}
F. Gross, J.~W.~Van~Orden and K.~Holinde,
%``Relativistic effects in low energy nucleon nucleon
%scattering,''
Phys.\ Rev.\ C {\bf 41}, R1909 (1990);
 F. Gross, J. W. Van Orden and K. Holinde,
%``Relativistic one boson exchange model for the nucleon
%nucleon interaction,''
 Phys.\ Rev. \ C {\bf 45}, 2094
(1992).



\bibitem{said}
R.~A.~Arndt, W.~J.~Briscoe, I.~I.~Strakovsky and R.~L.~Workman,
  %``Updated analysis of NN elastic scattering to 3 GeV,''
  Phys.\ Rev.\  C {\bf 76}, 025209 (2007)
  [arXiv:0706.2195 [nucl-th]]; data available through
SAID, http://gwdac.phys.gwu.edu/



\bibitem{misak}
M.~M.~Sargsian,
  %``Selected topics in high energy semi-exclusive electro-nuclear  reactions,''
  Int.\ J.\ Mod.\ Phys.\  E {\bf 10}, 405 (2001)
  [arXiv:nucl-th/0110053];
  %%CITATION = IMPAE,E10,405;%%
M.~M.~Sargsian, T.~V.~Abrahamyan, M.~I.~Strikman and
L.~L.~Frankfurt,
  %``Exclusive electrodisintegration of He-3 at high Q**2. I. Generalized
  %eikonal approximation,''
  Phys.\ Rev.\  C {\bf 71}, 044614 (2005)
  [arXiv:nucl-th/0406020];
  L.~L.~Frankfurt, M.~M.~Sargsian and M.~I.~Strikman,
  %``Feynman graphs and generalized eikonal approach to high energy  knock-out
  %processes,''
  Phys.\ Rev.\  C {\bf 56}, 1124 (1997)
  [arXiv:nucl-th/9603018].

\bibitem{genteikonal}
  J.~Ryckebusch, D.~Debruyne, P.~Lava, S.~Janssen, B.~Van Overmeire and T.~Van Cauteren,
  %``Relativistic formulation of Glauber theory for A(e,e' p) reactions,''
  Nucl.\ Phys.\  A {\bf 728}, 226 (2003)
  [arXiv:nucl-th/0305066];
  %%CITATION = NUPHA,A728,226;%%
D.~Debruyne, J.~Ryckebusch, W.~Van Nespen and S.~Janssen,
  %``The relativistic eikonal approximation in high-energy A(e,e' p)
  %reactions,''
  Phys.\ Rev.\  C {\bf 62}, 024611 (2000)
  [arXiv:nucl-th/0005058];
  %%CITATION = PHRVA,C62,024611;%%
  B.~Van Overmeire and J.~Ryckebusch,
  %``Second-Order Eikonal Corrections for A(e,e'p),''
  Phys.\ Lett.\  B {\bf 650}, 337 (2007)
  [arXiv:0704.0705 [nucl-th]];
  %%CITATION = PHLTA,B650,337;%%
W.~Cosyn and J.~Ryckebusch,
  %``On the density dependence of single-proton and two-proton knockout
  %reactions under quasifree conditions,''
  arXiv:0904.0914 [nucl-th].



\bibitem{ciofi}
C.~Ciofi~degli Atti and L.~P.~Kaptari,
  %``Calculations of the exclusive processes **2H(e,e**'p)n,**3He(e,e**'p)**2H,
  %and **3He(e,e**'p)(pn) within a generalized eikonal approximation,''
  Phys.\ Rev.\  C {\bf 71}, 024005 (2005);
  %%CITATION = PHRVA,C71,024005;%%
C.~Ciofi delgi Atti and L.~P.~Kaptari,
  %``A non factorized calculation of the process ^3He(e,e'p)^2H at medium
  %energies,''
  arXiv:0705.3951 [nucl-th];
  %%CITATION = ARXIV:0705.3951;%%
C.~Ciofi degli Atti, L.~P.~Kaptari and D.~Treleani,
  %``On the effects of the final state interaction in the
  %electro-disintegration of the deuteron at intermediate and high  energies,''
  Phys.\ Rev.\  C {\bf 63}, 044601 (2001)
  [arXiv:nucl-th/0005027].
  %%CITATION = PHRVA,C63,044601;%%


\bibitem{laget}
  J.~M.~Laget,
  %``The electro-disintegration of few body systems revisited,''
  Phys.\ Lett.\  B {\bf 609}, 49 (2005)
  [arXiv:nucl-th/0407072].

\bibitem{schiavilla}
  R.~Schiavilla, O.~Benhar, A.~Kievsky, L.~E.~Marcucci and M.~Viviani,
  %``Two-body electrodisintegration of He-3 at high momentum transfer,''
  Phys. Rev.  C{\bf 72}, 064003 (2005)
  [arXiv:nucl-th/0508048].
  %%CITATION = PHRVA,C72,064003;%%



\bibitem{halladata}
Jefferson Lab Experiment E01 - 020, spokespersons W. Boeglin, M. Jones, A. Klein, P. Ulmer,
J. Mitchell, E. Voutier.

\bibitem{egiyansrc}
K.~S.~Egiyan {\it et al.}  [CLAS Collaboration],
  %``Measurement of 2- and 3-Nucleon Short Range Correlation Probabilities in
  %Nuclei,''
  Phys.\ Rev.\ Lett.\  {\bf 96}, 082501 (2006)
  [arXiv:nucl-ex/0508026];
  %%CITATION = PRLTA,96,082501;%%
K.~S.~Egiyan {\it et al.}  [CLAS Collaboration],
  %``Observation of Nuclear Scaling in the $A(e,e^{\prime})$ Reaction at
  %$x_B>$1,''
  Phys.\ Rev.\  C {\bf 68}, 014313 (2003)
  [arXiv:nucl-ex/0301008].

\bibitem{blast}
BLAST data from MIT Bates, Ph. D. thesis A. Maschinot (MIT 2005).


\bibitem{jerrygexp}
G.~Gilfoyle, spokesperson, Jefferson Lab Hall B, E5 run period;
G.P. Gilfoyle, (the CLAS Collaboration),
'Out-of-Plane Measurements of the Fifth Structure Function of the Deuteron',
Bull. Am. Phys. Soc., Fall DNP Meeting, DF.00010(2006).


\bibitem{wernernewprop}
W. Boeglin, spokesperson, proposal to Jefferson Lab PAC 33, 2007.

\bibitem{sebastian}
Jefferson Lab 93-009 (EG1 run group), G. Dodge, S. Kuhn and M. Taiuti, co-spokespersons.

\bibitem{doug}
I. Passchier et al., Phys.\ Rev.\ Lett.\ {\bf 88}, 102302,2002.



\bibitem{blast2}
  E.~Geis {\it et al.}  [BLAST Collaboration],
  %``The Charge Form Factor of the Neutron at Low Momentum Transfer from the
  %$^{2}\vec{\rm H}(\vec{\rm e},{\rm e}'{\rm n}){\rm p}$ Reaction,''
  Phys.\ Rev.\ Lett.\  {\bf 101}, 042501 (2008)
  [arXiv:0803.3827 [nucl-ex]].


\bibitem{arenhovel}
  H.~Arenhovel, W.~Leidemann and E.~L.~Tomusiak,
  %``THE ROLE OF THE NEUTRON ELECTRIC FORM-FACTOR IN D (E, E-PRIME N) N
  %INCLUDING POLARIZATION OBSERVABLES,''
  Z.\ Phys.\  A {\bf 331}, 123 (1988);
  H.~Arenhoevel, W.~Leidemann and E.~L.~Tomusiak,
  %``Exclusive Deuteron Electrodisintegration With Polarized Electrons And A
  %Polarized Target,''
  Phys.\ Rev.\  C {\bf 46}, 455 (1992).

\bibitem{sabinetensor}
  A.~Bianconi, S.~Jeschonnek, N.~N.~Nikolaev and B.~G.~Zakharov,
  %``Quadrupole deformation of deuterons and final state interactions in H-2
  %(polarized) $\to$ (e, e-prime p) scattering on tensor polarized deuterons at
  %CEBAF energies,''
  Phys.\ Rev.\  C {\bf 53}, 576 (1996)
  [arXiv:nucl-th/9501017].


\bibitem{boeglintrento2005}
Werner Boeglin, talk at the 2005 Workshop on ``Probing microscopic
structure of the lightest nuclei in electron scattering at JLab
energies and beyond'', Trento, Italy July 25-30, 2005,
http://www.fiu.edu/~sargsian/ect05



\bibitem{raskintwd}
A. S. Raskin and T. W. Donnelly, {\sl Ann. of Phys.} {\bf 191}, 78
(1989).


\bibitem{dmtrgross}
V. Dmitrasinovic and F. Gross, {\sl Phys. Rev. C} {\bf40}, 2479
(1989).


\end{thebibliography}
\end{document}